\documentclass[journal]{IEEEtran}

\ifCLASSINFOpdf
\usepackage{soul}
\usepackage[pdftex]{graphicx}
\usepackage{amssymb}
\usepackage{amsmath}
\usepackage{gensymb}
\usepackage{xcolor}
\usepackage{array}
\usepackage{float}
\usepackage{subcaption,graphicx}
\usepackage{xspace}
 \else
 
\fi

\usepackage{array}
\usepackage{threeparttable}
\usepackage{tablefootnote}

\begin{document}
%
\title{Towards Indirect Data-Driven Predictive Control for Heating Phase of Thermoforming Process}
%
%
\newcommand{\MATLAB}{\textsc{Matlab}\xspace}

\author{Hadi Hosseinionari, Mohammad Bajelani, Klaske van Heusden, Abbas S. Milani,  and Rudolf Seethaler
\thanks{We acknowledge the support of the New Frontiers in Research Fund (NFRF) of Canada, [NFRFE-2019-01440] and the Natural Sciences and Engineering Research Council of Canada (NSERC) [RGPIN-2023-03660].}
\thanks{Hadi Hosseinionari, Mohammad Bajelani, Klaske van Heusden, Abbas S. Milani,  and Rudolf Seethaler are with the University of British Columbia, School of Engineering, 1137 Alumni Avenue, Kelowna, BC V1V 1V7
{\tt\small hadi.hosseinionari, mohammad.bajelani, klaske.vanheusden, abbas.milani, rudolf.seethaler @ubc.ca}}}



\markboth{}
{Shell \MakeLowercase{\textit{et al.}}: Bare Demo of IEEEtran.cls for IEEE Journals}

\maketitle

\begin{abstract}

Shaping thermoplastic sheets into three-dimensional products is challenging since overheating results in failed manufactured parts and wasted material. To this end, we propose an indirect data-driven predictive control approach using Model Predictive Control (MPC) capable of handling temperature constraints and heating-power saturation while delivering enhanced precision, overshoot control, and settling times compared to state-of-the-art methods. We employ a Non-linear Auto-Regressive with Exogenous inputs (NARX) model to define a linear control-oriented model at each operating point. Using a high-fidelity simulator, several simulation studies have been conducted to evaluate the proposed method's robustness and performance under parametric uncertainty, indicating overshoot and average steady-state error less than $2^\circ \mathrm{C}$ and $0.7^\circ \mathrm{C}$ ($7^\circ \mathrm{C}$ and $2^\circ \mathrm{C}$) for the nominal (worst-case) scenario. Finally, we applied the proposed method to a lab-scale thermoforming platform, resulting in a close response to the simulation analysis with overshoot and average steady-state error metrics less than $5.3^\circ \mathrm{C}$ and $1^\circ \mathrm{C}$, respectively.

\end{abstract}


\IEEEpeerreviewmaketitle

\section{Introduction}
\label{sec: Introduction}

\begin{figure}[htpb]
    \centering
    \begin{subfigure}[b]{0.5\textwidth}
        \centering
        \includegraphics[width=\textwidth]{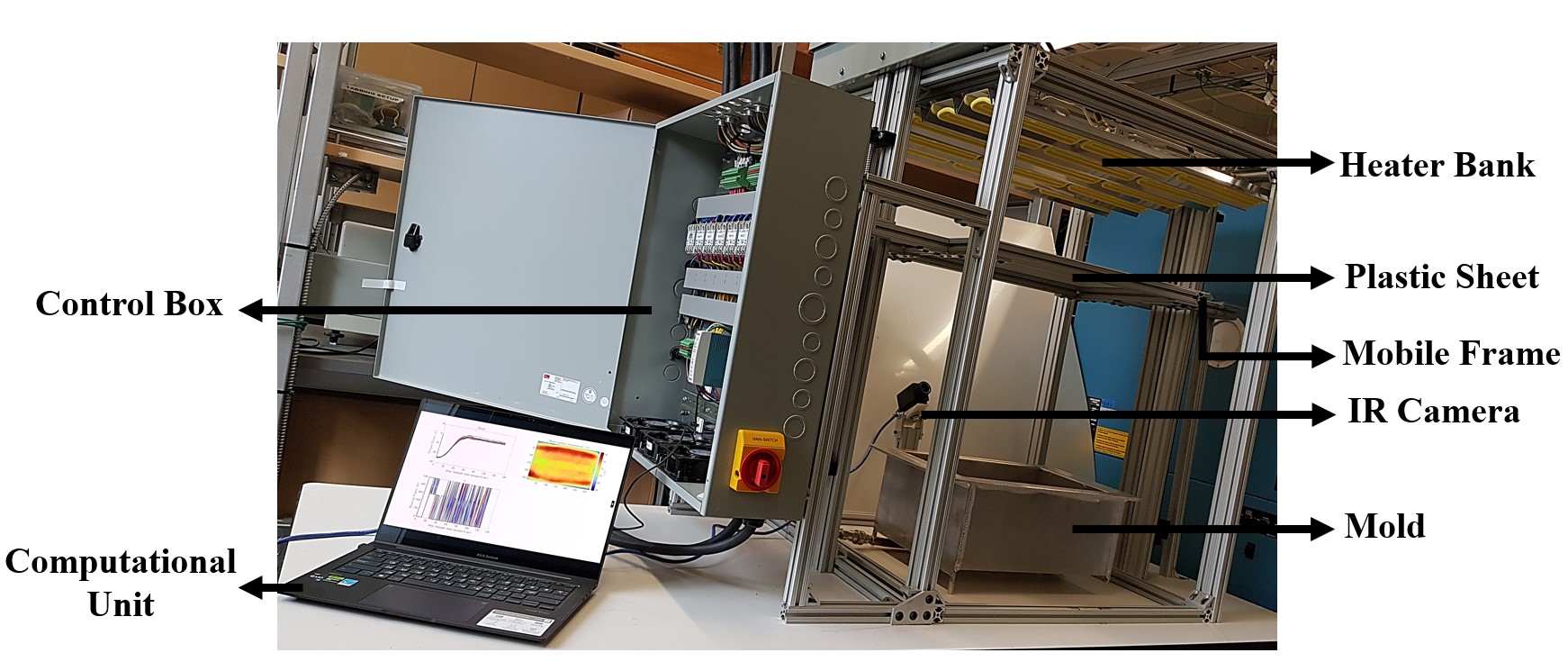}
        \caption{Lab-scale thermoforming setup}
        \label{fig: setup}
    \end{subfigure}
    \begin{subfigure}[b]{0.4\textwidth}
        \centering
        \includegraphics[width=\textwidth]{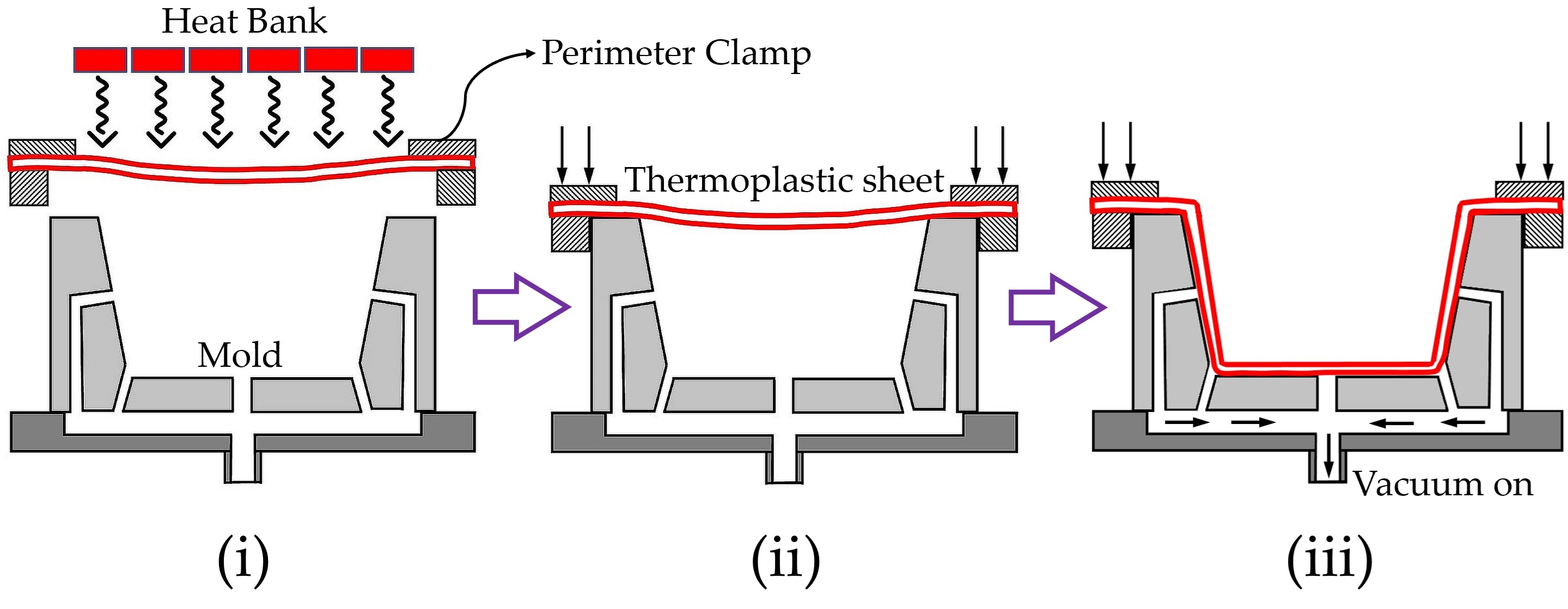}
        \caption{Main sequences of a thermoforming process: (i) heating (ii) sealing or pre-stretch (iii) forming and cooling \cite{hosseinionari2023development}.}
        \label{fig: Thermoforming Process}
    \end{subfigure}
    \caption{An overview of a typical thermoforming process employed in this study.}
    \label{fig: overview}
\end{figure}

Composite manufacturing has received considerable attention in Industry 5.0 \cite{xu2021industry, leng2022industry}. More specifically, thermoplastic composites are increasingly used in the aerospace, marine, renewables, and automotive industries because of their thermo-mechanical properties and lightweight characteristics \cite{throne2013technology}. A thermoforming process involves applying heat and pressure to reshape flat plastic sheets, resulting in three-dimensional products. Fig. \ref{fig: overview} shows a thermoforming setup with process sequences. The overall quality of the thermoformed product mainly depends on the heating phase's accuracy aligned with the mold geometry \cite{tam2007thermoforming}, i.e., temperature deviations can potentially result in undesired wrinkles and thickness \cite{boisse2016modelling}. Temperature control effectively helps reduce the number of rejected parts while minimizing cycle times and lowering the production cost. The main objective of this paper is to address the heating phase control problem using data-driven techniques.

Employed methods in the thermoforming industry are limited in rule-based control laws requiring expert knowledge or basic control strategies \cite{Albadawi2006}. Traditional control methods cannot accommodate process constraints, including maximum and minimum temperature and heating/cooling rates \cite{Borrelli_Bemporad_Morari_2017}. In \cite{li2010study}, the optimal heater input over time is investigated and subsequently established to reduce the temperature disparity across the thickness for uniform temperature distribution of the sheet. In \cite{Chy2011MPC}, a physics-based Model Predictive Control (MPC) model designed for the heating phase of the thermoforming is proposed, resulting in a high-computational cost, required full-state feedback solution and only evaluated using simulation studies for the nominal case. In \cite{HOSSEINIONARI2024DRL}, an approach combining MPC and Deep Reinforcement Learning (DRL) is introduced to enhance the efficiency of radiation thermal control systems. The main advantage of this method is that the trained agent performs similarly to Adaptive Model Predictive Control (AMPC) while having a low online computational load. The drawback of this method is that the offline training process increases with the number of heaters.

MPC's most challenging and time-consuming component is the prediction model \cite{richalet1993industrial, zhu1998multivariable, ZHU2013MPC}. Typically, this model can be gained from physics rules or System Identification (SysID) techniques. Available industrial datasets help us to adopt novel data-driven predictive-based control approaches capable of handling constraints \cite{Chy2011MPC,dutta2014certification}. The present paper incorporates a control-oriented model derived from a NARX model into the MPC framework. This method enables us to handle complex nonlinear systems with input-output constraints that suit the thermoforming control problem.

In manufacturing, applying NARX-MPC has demonstrated significant potential in stabilizing systems under various conditions. In scenarios where the system dynamics are not fully known but input-output data is readily available, applying NARX-MPC has demonstrated significant potential in stabilizing systems under various conditions \cite{seel2021neural}. Furthermore, the ability of NARX-MPC to adapt to varying operational conditions with quicker response times and reduced errors has proven crucial in maintaining optimal product concentration and desired temperature \cite{mohd2015control}. In \cite{NIKOLAKOPOULOU2023108272}, NARX-MPC has been applied in chemical processes to control a continuous stirred-tank reactor with a heated jacket. The methodology proposed for identifying sparse polynomial NARX models using elastic net for real-time implementable MPC demonstrated satisfactory closed-loop performance under various setpoints, unmeasured disturbances, and measurement noise. In smart buildings, a learning-based MPC approach for thermal control has been proposed, utilizing NARX models to optimize the thermal environment efficiently in \cite{Eini2019Learning}. In \cite{HU2024NARX}, an accurate combustion process prediction model based on NARX is established to deploy with an MPC control strategy and realize the uniform combustion in the boiler by controlling the opening travel of secondary windgates. In \cite{JUNG2023108447}, NARX-MPC is applied on the post-combustion $\mathrm{CO}_2$ capture (PCC) process to balance model complexity with prediction accuracy. In \cite{bonassi2022offset}, a method is proposed to design a control strategy based on NARX models that ensure offset-free tracking of constant references while maintaining the nominal stability of the closed-loop system. NARX-MPC has also been adopted for industrial applications like bioethanol fermentation process control\cite{LEE2018NARX-GAS}. All these methods show the feasibility of adopting the NARX-MPC strategy for manufacturing settings like thermoforming systems.

In this paper, we adopt a linearized version of the NARX model at each operating point, resulting in a computationally efficient and easy-to-tune approach. We aim to handle the MIMO-constrained thermoforming system by designing an MPC to achieve a high-performance control strategy. The main component of MPC is the control-oriented model. Generally, this model must capture the system's essential dynamics while being computationally feasible. We use a high-fidelity simulator to gather a dataset in various conditions to save time and energy. Then, we perform a data-driven model reduction to find a simple linear representation that can be used in linear MPC. This indicates that, in practice, identifying a low-order model results in high-performance control and simpler implementation. This paper's main contribution is to develop and implement an indirect data-driven control method, resulting in a robust solution to sensitive parameters and accurate solutions compared to two state-of-the-art methods, Adaptive MPC (AMPC) and MPC-guided Deep Reinforcement Learning (DRL) for the processes  \cite{HOSSEINIONARI2024DRL}.

The subsequent sections of this paper are structured as follows: Section \ref{sec: Modeling} briefly explains the modeling and heat transfer equations. Section \ref{sec: Experimental Setup} provides a detailed overview of the employed thermoforming experimental setup. Moving on to Section \ref{Data-Driven Modeling and Control}, we elucidate the proposed indirect data-driven predictive-based method design. Section \ref{Proposed Control Strategy for Thermoforming Systems} specifies the proposed control strategy for the thermoforming system. Next, section \ref{Results} compares results obtained from the simulator and the real-world setup. Lastly, Section \ref{sec: Conclusion} encapsulates the conclusions drawn from the study and outlines potential avenues for future research.

\section{Heat Transfer Model}
\label{sec: Modeling}

\begin{figure} [!]
  \centering
      \includegraphics[width=0.45\textwidth]{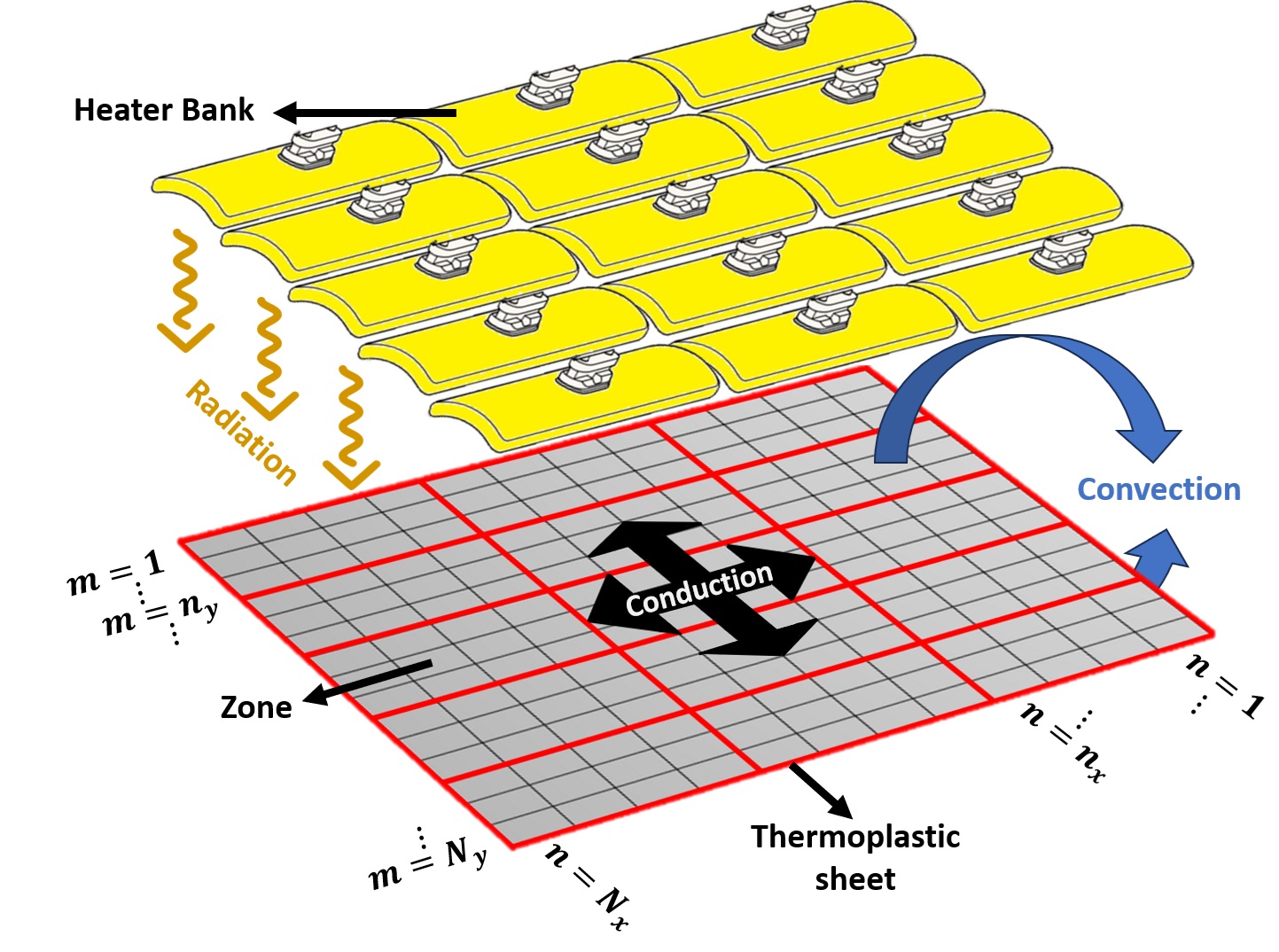}
      \caption{Thermoplastic sheet meshing and definition of the zone.}      \label{fig: zone} 
\end{figure}

The heat transfer model employed in this study follows the methodology outlined in the authors' previous work, which simulated the dynamics of a laboratory-scale thermoforming system. A brief overview is provided here, while more comprehensive details can be found in \cite{hosseinionari2023development}. During the thermoforming process's heating phase, the primary heat transfer mode is from the heater bank to the sheet's surface through radiation. Depending on the material properties of the thermoplastic sheet, a fraction of the Infrared (IR) radiation is absorbed, leading to an increase in the temperature of the sheet. The thermoplastic sheet's elevated temperature results in heat dissipation to the environment through both convection and radiation. Additionally, assuming that the thermoplastic sheet is divided into $N_x \times N_y$ equal elements according to Fig. \ref{fig: zone}, conduction heat transfer occurs between adjacent elements on the thermoplastic sheet. The heat balance equation for each element on the sheet can be expressed as follows:

\begin{equation}
\rho V c_p \frac{\partial T_{i}}{\partial t} = \nabla \cdot (\overrightarrow{Q_\mathrm{rad}})_{i} + \nabla \cdot (\overrightarrow{Q_\mathrm{conv}})_{i} + \nabla \cdot (\overrightarrow{Q_\mathrm{cond}})_{i}
\label{eq: balance}
\end{equation}

In this equation, $T_{i}$ is the temperature of $i$th element on the thermoplastic sheet. $Q_\mathrm{rad}$, $Q_\mathrm{conv}$, and $Q_\mathrm{cond}$ are radiation, convection, and conduction heat fluxes, respectively, $\mathrm{\rho}$ are sheet material's density, $V$ is the sheet element's volume, and $\mathrm{c_p}$ is specific heat. Radiation from the heater bank is the primary heat flux source toward elements on the thermoplastic sheet. The following equation represents this radiation heat flux.
\begin{equation}
 Q_{\mathrm{rad}} = {A_h \varepsilon_e \sigma} \sum_{i=1}^{N_xN_y} \sum_{h=1}^{H} F_{h \to i} (\theta_h^4 - T_i^4),
\label{eq: radiation}
\end{equation}
where $A_h$ is the surface area of the heaters, $\epsilon_e$ shows the effective emissivity from heater to sheet, $\sigma$ indicates the Stefan-Boltzmann constant, $H$ is the number of heaters in the heater bank, $F_{h \to i}$ is the view factor between the $h^{th}$ heater in the heater bank and the $i^{th}$ element on the thermoplastic sheet, $\theta_h$ is the surface absolute temperature of the $h^{th}$ heater in the heater bank, and $T_i$ is the absolute temperature of the $i^{th}$ element on the thermoplastic sheet. Heat dissipation to the surroundings occurs through radiation and convection heat transfer from the sheet elements. Notably, the convection heat transfer coefficient varies between the top and bottom faces of the sheet. Equation \eqref{eq: Convection} delineates the convection heat flux from a sheet element to the environment.

\begin{equation}
 Q_{\mathrm{conv}} = (h_t+h_b) \Delta x \Delta y \Delta T,
\label{eq: Convection}
\end{equation}
where $h_t$ and $h_b$ represent the convection heat transfer coefficients at the top and bottom faces of the sheet, respectively, $\Delta x \Delta y$ denotes the surface area of the sheet element exposed to the air, and $\Delta T$ is the temperature difference between the sheet element and the ambient. Additionally, heat transfer through conduction occurs in adjacent elements from elements with higher temperatures to those with lower temperatures on the thermoplastic sheet. However, this type of heat transfer is relatively small compared to radiation and convection, owing to the low thermal conductivity of thermoplastic materials. Equation \eqref{eq: conduction} details the conduction heat flux between sheet elements.
\begin{equation}
 Q_{\mathrm{cond}} = \mathrm{k} A_e \frac{\Delta T}{\Delta L},
\label{eq: conduction}
\end{equation}
where k is the thermoplastic sheet material thermal conductivity, $A_e$ is the element's cross-sectional surface area, $\Delta L$ is the width of the element in the $\mathrm{x}$ or $\mathrm{y}$ directions, and $\Delta T$ is the temperature difference between the ends. Figure \ref{fig: zone} illustrates the heater bank alongside the thermoplastic sheet, partitioned into $N_x \times N_y$ equal elements. The figure delineates three types of heat transfer transpiring during the thermoforming heating phase. The region beneath each heating element on the thermoplastic sheet is called a ``Zone". Every zone receives radiation from all heaters in the heater bank; however, the predominant infrared radiation received is from the heater directly above that specific zone. In the scenario where the sheet is discretized into cubic elements referred to as control volumes, equation (\ref{eq: balance}) is integrated over each control volume and across the time interval from $t + \Delta t$. Hence, the temperature change in each sheet element is determined by the change in its internal energy, as expressed in equation \eqref{eq: T_t}.

\begin{equation}
 T_{t+\Delta t} = T_{t} +( \frac{\Delta t}{\rho V c_p}) dU,
\label{eq: T_t}
\end{equation}
where $dU=Q_\mathrm{rad} + Q_\mathrm{conv} + Q_\mathrm{cond}$.

\section{Experimental Setup}
\label{sec: Experimental Setup}

The physical setup illustrated in Fig. \ref{fig: setup} is a lab-scale thermoforming system with fifteen ceramic heating elements (FTE-500, Ceramicx Company) arranged in five rows and three columns in the heater bank. Each zone receives radiation from all heaters in the heater bank, indicating a Multi-Input Multi-Output (MIMO) system. A mobile frame holds the thermoplastic sheet during the heating phase and transfers it to the mold once the heating phase is completed. A FLIR Infrared (IR) camera (FLIR A35) provides temperature feedback from the thermoplastic sheet's surface to the control algorithm with a 60 Hz frame rate and 16-bit video streaming. The compact dimensions of the thermal camera (29 × 36 × 59 mm) enhance its versatility within the experimental setup. Control input values are transmitted to an Arduino Atmega 2560 via a serial port, where they are processed to manage the states of the relays (Omega - SSRL240AC10). The input values, derived from MATLAB and representing the output of the control algorithm within the range of 0 to 255 units, are used by the Arduino firmware to generate PWM signals. The AC unit is segregated into two circuits to manage the power requirements. The first circuit powers eight heaters, each drawing 2.5 Amps, while the second circuit caters to seven heaters with the same power draw. Each unit operates within a range of 220V and up to 20 Amps. Table \ref{tab: system parameters} shows this study's model parameters.

\newcolumntype{P}[1]{>{\centering\arraybackslash}p{#1}}
\begin{table}[!]
    \centering
    \small
    \caption{System parameters.}
    \label{tab: system parameters}
    \begin{tabular}{P{1cm} P{3.5cm} P{1cm} P{1cm}}
    \hline
    \textbf{Parameter} & \textbf{Description} & \textbf{Values} & \textbf{Unit}\\
    \hline
$\rho$	& sheet density &1380& $[\frac{\mathrm{kg}}{\mathrm{m}^3}]$\\
$\mathrm{c}_\mathrm{p}$ &  specific heat capacity &	1465& $\left[\frac{\mathrm{J}}{\mathrm{kg} \cdot \mathrm{K}}\right]
$\\
$\epsilon_e$& effective emissivity &	0.95& -\\
$\mathrm{k}$&  material thermal conductivity &	0.18& $\left[\frac{\mathrm{W}}{\mathrm{m} \cdot \mathrm{K}}\right]$\\
$\mathrm{h}$& convection heat transfer coefficient &5& $[ \frac{\mathrm{W}}{ \mathrm{m}^2 \cdot \mathrm{K}}]$\\
$\Delta z$	& sheet thickness & 0.002& $[ \mathrm{m}]$\\
$\mathrm{L}_x$ & sheet length&	0.75& $[ \mathrm{m}]$\\
$\mathrm{L}_y$	& sheet width &0.5& $[ \mathrm{m}]$\\
    \hline
    \end{tabular}
\end{table}

\section{Data-Driven System Modeling and Control}
\label{Data-Driven Modeling and Control}


This section introduces the employed nonlinear system identification technique, its linearized representation, and an indirect data-driven predictive control approach.
\subsection{Data-Driven System Modeling: Nonlinear Autoregressive Exogenous (NARX)}

\begin{figure}[!]
  \centering
      \includegraphics[width=0.4\textwidth]{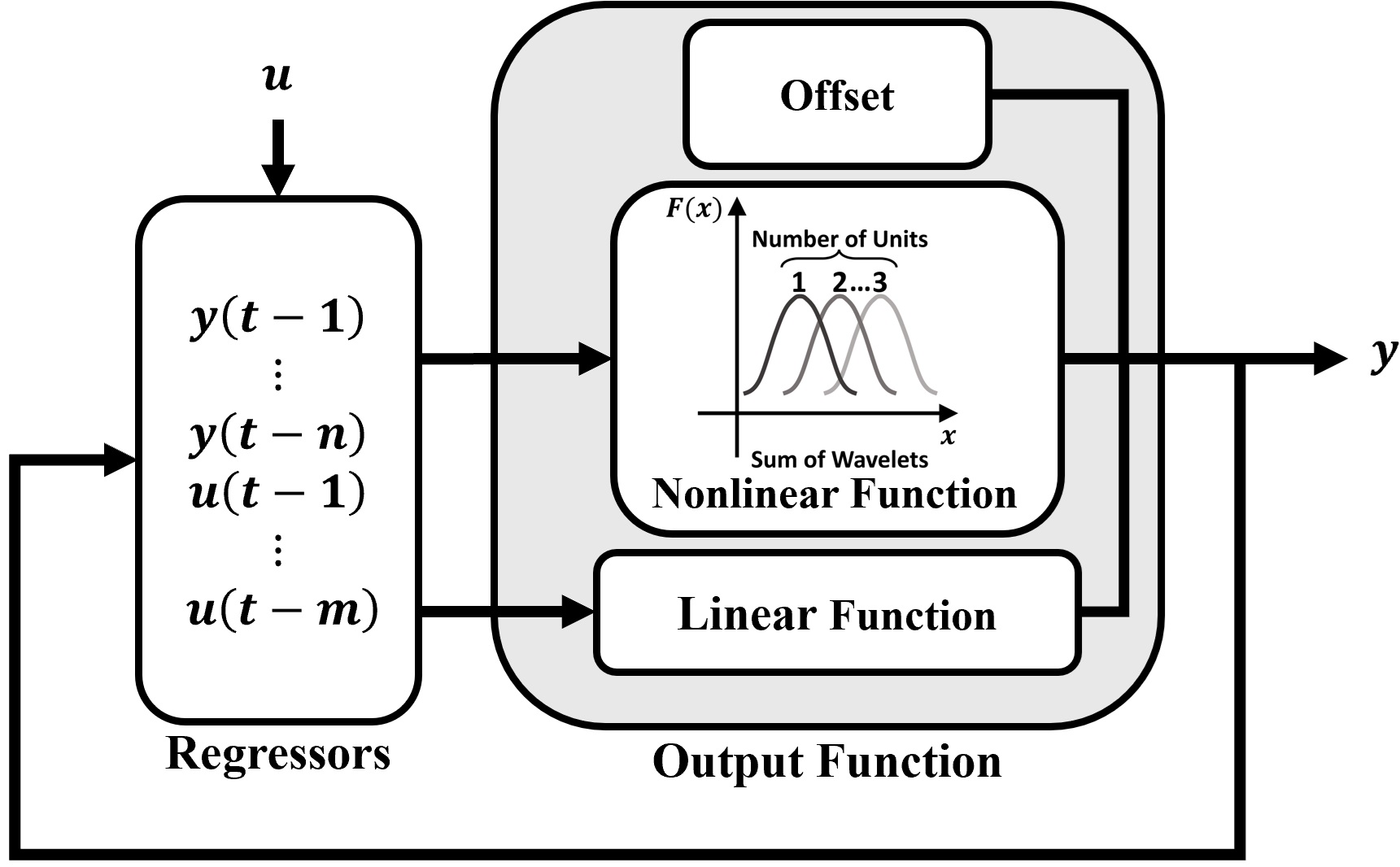}
      \caption{ A single-output NARX model block diagram.}  \label{fig: NARX Flowchart} 
\end{figure}


A NARX model extends the traditional linear ARX structure by incorporating linear and nonlinear components in parallel; see Fig. \ref{fig: NARX Flowchart}. It consists of two main components: a regressor vector and an output function. The regressor vector contains past input-output data, indicating the model's order. The output function involves offset, nonlinear, and linear functions, indicating the model's structure. Since the output function benefits from a parallel structure, its nonlinear and offset parts are dedicated to reducing the modeling error that cannot be captured by the linear part. A NARX model with a wavelet nonlinear function can be represented as follows:
\begin{subequations}\label{equ:NARX-regressors}
\begin{align}
    X_t &= \left[\begin{array}{l}
    y_{[t-1:t-n]}, u_{[t-1:t-m]} 
    \end{array}\right]^{\top}, \label{equ:NARX-general1} \\
    y_t(X_t) &= X_t^T P L + W(X_t) + S(X_t) + \varepsilon, 
\end{align}
\end{subequations}
where $ X_t \in \mathbb{R}^{(n+m)\times1}$ the regressor vector at time step $ t $, $ \varepsilon $ is the offset, $ P \in \mathbb{R}^{(n+m)\times p}$ is a projection matrix, transforming the regressors into a lower-dimensional space, $ L \in \mathbb{R}^{p\times1}$ is linear mapping that scales the projected regressors to the output. $ W(X_t) $ represents the sum of dilated and translated wavelets, which are functions that capture the localized features of the signal. $ S(X_t) $ represents the sum of dilated and translated scaling functions, also known as scales, which capture the global trends of the signal. The functions $ W(X_t) $ and $ S(X_t) $ together form the wavelet network's nonlinear component, allowing the NARX model to approximate complex, nonlinear relationships between the inputs and outputs \cite{zhang1997wavelet}. To linearize the NARX model, consider the following equation:
\begin{subequations}\label{equ:Linearization}
\begin{align}
    X_{t+1} &= AX_{t} + [B_1 \ B_2] \begin{bmatrix}
    y_t \\
    u_t
    \end{bmatrix}, \label{equ:Linearization-1} \\
    y_t &= \mathcal{F}(X_t, u_t) \label{equ:Linearization-2}
\end{align}
\end{subequations}
where $\mathcal{F}$ is a nonlinear function representing the output function in Figure \ref{fig: NARX Flowchart}. $X_t$ and $ u_t $ are state and the input vectors at time $t$, respectively. To establish a linear state-space model around the operating point $y_{op} = \mathcal{F}(X_{op}, u_{op})$, the first derivatives of the function $ \mathcal{F} $ with respect to the state vector $ X $ and the input $ u $ are represented as follows:
\begin{equation} \label{equ:Linearization-4}
    \mathcal{F}_X = \frac{\partial}{\partial X} \mathcal{F}(X, u) \bigg|_{\substack{u = u_{op} \\ X = X_{op}}},
    \quad \mathcal{F}_u = \frac{\partial}{\partial u} \mathcal{F}(X, u) \bigg|_{\substack{u = u_{op} \\ X = X_{op}}},
\end{equation}
Using the derivatives, the changes in the output and state vector from their values at the operating point can be approximated by the following linear equations:
\begin{subequations}\label{equ:Linearization-2}
\begin{align}
    \Delta X_{t+1} &= A \Delta X_t + [B_1 \ B_2] \begin{bmatrix}
    \Delta y_t \\
    \Delta u_t
    \end{bmatrix}. \label{equ: Linearization-5}\\
    \Delta y_t &= \mathcal{F}_X \Delta X_t + \mathcal{F}_u \Delta u_t. \label{equ: Linearization-6}
\end{align}
\end{subequations}
By substituting equation \eqref{equ: Linearization-6} in \eqref{equ: Linearization-5}, the linear approximation for the state vector around the operational point is defined as follows:
\begin{equation} \label{equ:Linearization-7}
\Delta X_{t+1} = (A + B_1 \mathcal{F}_X) \Delta X_t + (B_1 \mathcal{F}_u + B_2) \Delta u_t
\end{equation}
where $\Delta X_t = X_t - X_{op}$, $\Delta u_t = u_t - u_{op}$, $\Delta y_t = y_t - y_{op}$ are the deviation of the state vector, input, and output from their value at the operating point, respectively. $B_1$ and $B_2$ are matrices that relate the input and the output to the state changes. This method is implemented by the ``linearize'' command in MATLAB\textsuperscript{\textregistered} - System Identification Toolbox.

\subsection{Model Predictive Control (MPC)}

\begin{figure}[!]
  \centering
      \includegraphics[width=0.4\textwidth]{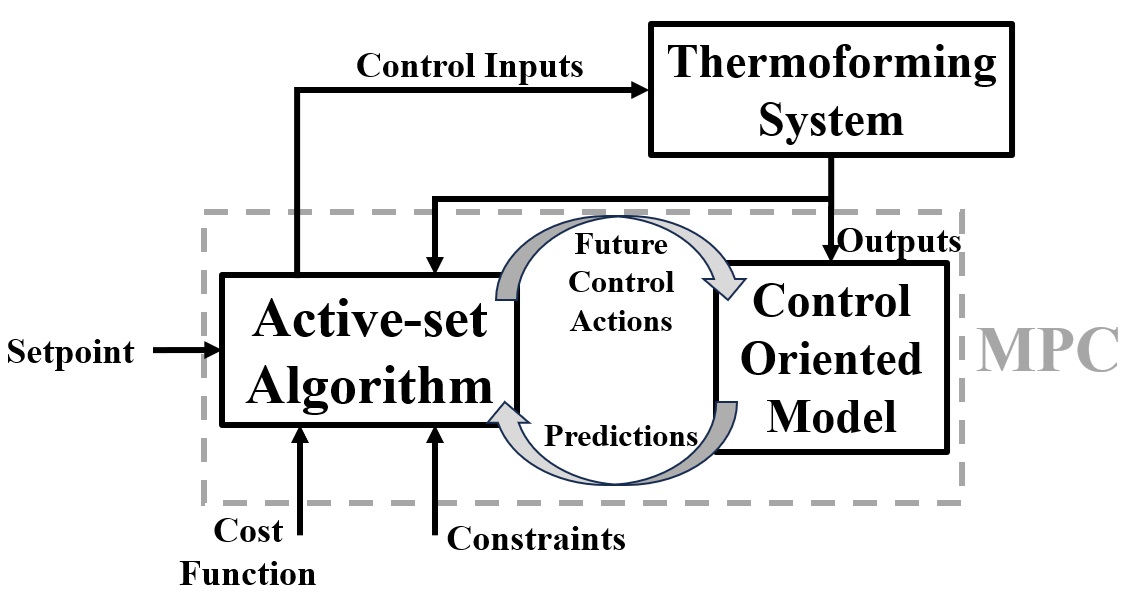}
      \caption{ MPC block diagram \cite{HOSSEINIONARI2024DRL}.}   
      \label{fig: MPC general} 
\end{figure}

MPC is an advanced control technique that enables us to predict a system's trajectory over a finite prediction horizon with the cost of solving an on-the-fly optimization problem. As an optimization-based controller, MPC can systematically handle input-output constraints \cite{rawlings2017model}. The main bottleneck of this method is to achieve a \emph{control-oriented} model. Often, highly accurate and over-complicated models may not be suitable for control design as the underlying optimization problem becomes expensive and time-consuming, especially for high-dimensional systems. Therefore, we adopt the linearized NARX model, given in (\ref{equ: Linearization-5}-\ref{equ: Linearization-6}). For a servo-tracking problem\footnote{The reference, $r_t$, is a constant signal over the prediction horizon.}, we define the MPC problem as follows:

\begin{equation} \label{equ: MPC general}
\begin{array}{cl}
\underset{u}{\operatorname{minimize}} & \sum_{t=1}^{N_p}  \alpha  e_t^T e_t + \sum_{t=0}^{N_c-1} \beta \Delta u_t^T \Delta u_t  \\
\text { subject to } & X_{t+1}=\mathcal{A} X_t+\mathcal{B} u_t,\\
& y_{t}=\mathcal{C} X_t+\mathcal{D} u_t,\\
& e_t =  r - y_t ,\\
& y_t \in \mathcal{Y}, u_t \in \mathcal{U},\\
& 
\end{array}
\end{equation}
where $\alpha$ and $\beta$ penalize the tracking error and rate of the control inputs, $\mathcal{U}$ and $\mathcal{Y}$ are admissible sets for inputs and outputs, ensuring the system operates within safe and efficient bounds, $\mathcal{A}=A + B_1 \mathcal{F}_X$, $\mathcal{B}=B_1 \mathcal{F}_u + B_2$, $\mathcal{C}= \mathcal{F}_X$ and $\mathcal{D}=\mathcal{F}_u $ are the matrices of the control-oriented model from \eqref{equ: Linearization-5}. $N_p$ and $N_c$ are the prediction and control horizons, respectively. $\Delta u_t$ is the increment of control input at time step $t$. The general architecture of the proposed method is depicted in Fig. \ref{fig: MPC general}.


\section{Proposed Control Strategy for Thermoforming System} \label{Proposed Control Strategy for Thermoforming Systems}

This section specifies the proposed MPC method for the lab-scale thermoforming system, including the modeling phase, control objectives, and control law.

\subsection{Data-driven modeling for Thermoforming System}

\begin{figure*}[t]
    \centering
    \includegraphics[width=1\linewidth]{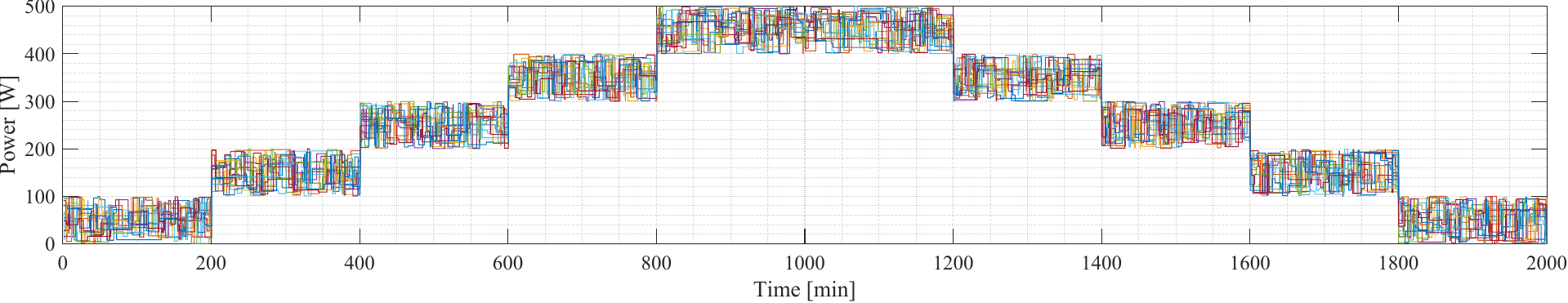}
    \caption{Proposed experiment design for data collection: excitation of input signals for 15 heaters (each colored curve refers to each heater).}
    \label{fig: u_excitation}
\end{figure*}

Experiment design\footnote{Data collection for SysID purposes.} for nonlinear systems requires a more profound intuition and physical understanding compared to linear systems, which primarily need to satisfy certain mathematical conditions such as persistently exciting (PE) inputs \cite{nelles2020nonlinear}. For a thermoforming system, we outline an experiment to collect data for modeling purposes with the following criteria: I) The input should drive the system across a broad range of sheet temperatures to expose the non-linearity of the system. II) The system's output should oscillate around various equilibrium points to capture the system's gain characteristics. III) Given the distinct nonlinear behaviors during thermoforming systems' warming and cooling phases, the input signal must be designed to explore these phases comprehensively. The system's non-linearity is gain-dependent and mainly depends on sheet temperature. We employ a pseudo-random binary sequence (PRBS) signal with the upper and lower bounds increasing every 200 minutes with the amplitude of $100 \, [\mathrm{W}]$ until reaching a maximum of $500 \, [\mathrm{W}]$, after which it decreases similarly; see Fig. \ref{fig: u_excitation}.

The optimal accuracy of the model is observed when parameters $n$ and $m$ in the regressor vector related to the number of past inputs and outputs are set to 2. It is carried out using MATLAB\textsuperscript{\textregistered} - System Identification Toolbox to achieve the best parameters and number of wavelets in the NARX model. 
\subsection{Control Objectives for Thermoforming Systems}

Our main objectives for controlling the heating phase of thermoforming systems are efficiency, reliability, and scalability.

\subsubsection{Heating beyond Glass Transition Temperature}
One of the critical control objectives is the handling of overheating (temperature overshoot). From a thermo-mechanical perspective, the appeal of amorphous thermoplastics lies in their wide forming window, which stems from their rubber-like behavior. This behavior occurs above the glass transition temperature $T_g$ and extends over various temperatures. The thermoformability of these materials hinges on finding a balance between withstanding significant deformations and maintaining a certain level of rigidity to prevent excessive sheet flow during heating. Typically, the forming window ranges from 30 to 60 $^\circ \mathrm{C}$ above $T_g$ for most amorphous thermoplastics \cite{schmidt1972biaxial}. The overshoot should be carefully controlled so the thermoforming processes do not exceed the forming window of the material. This constraint is essential to prevent material degradation and ensure product quality. Consequently, the admissible set, $\mathcal{Y}$ is defined as follows:
\begin{equation}
    \mathcal{Y} = \{\mathcal{T}_i | i\in \{1,...,15\} , \mathcal{T}_i-\mathcal{T}^{\, r}_i<\mathcal{T}_{over}\}
\end{equation}
where $\mathcal{T}_i$ and $\mathcal{T}^{\, r}_i$ are each zone's average and reference temperature on the thermoplastic sheet, respectively. $\mathcal{T}_{over}$ denotes the allowable overshoot temperature determined by the material properties, which is considered equal to zero in this paper.

\subsubsection{Operating Time Efficiency}
To enhance production efficiency, the control system must minimize the rise time and settling time during the heating phase. A rapid and accurate approach to the desired temperature setpoint, without excessive overshoot, is essential for increasing throughput. The proposed MPC's predictive capabilities allow for aggressive yet safe control actions. This can be achieved by setting a higher value for $\alpha$ compared to $\beta$ in equation \eqref{equ: MPC general}.

\subsubsection{Minimizing the terminal errors}
To maintain precise temperature control and ensure product quality, the control system must minimize the terminal error at the end of each control horizon. Specifically, the absolute difference between the desired set-point temperature and the actual temperature of each zone should be less than the required control accuracy, which is considered 5 \textdegree C in this study. This constraint can be expressed as:
\begin{equation}
|\mathcal{T}_i(t_f) - \mathcal{T}^{r}_i| < 5^\circ\mathrm{C}, \quad \forall i \in {1, \ldots, 15}
\end{equation}
where $t_f$ denotes the end of the control horizon, and $\mathcal{T}_i(t_f)$ represents the actual temperature of zone $i$ at time $t_f$. Enforcing this terminal error constraint ensures that the temperature distribution across the thermoplastic sheet is within an acceptable tolerance range, preventing material degradation and maintaining consistent product quality.

\subsubsection{Robustness}

\begin{figure}[!]
  \centering
      \includegraphics[width=0.35\textwidth]{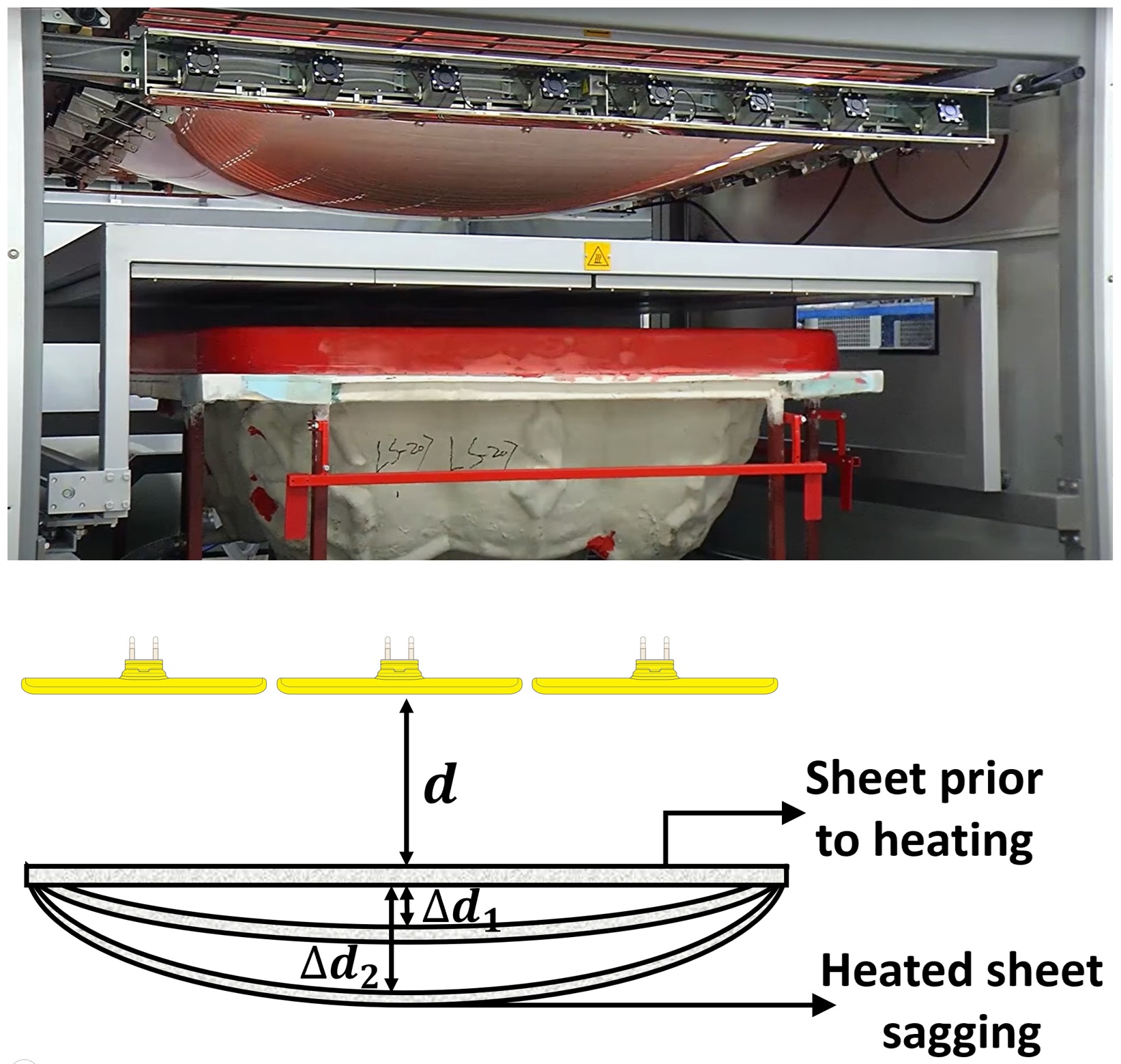}
      \caption{ An industrial example of large sheet sagging during the heating phase: CMS plastic technology company \cite{CMS}}   
      \label{fig: sag} 
\end{figure}

Industrial environments present numerous challenges, including variations in system parameters such as ambient temperature, which can affect the thermoforming process. The gap between the heater bank and the thermoplastic sheet is one of the most important system parameters. A smaller gap reduces disturbance from neighboring heaters, resulting in system behavior similar to a Single-Input Single-Output (SISO) configuration for each zone. In contrast, this minor gap may form hot spots on the thermoplastic sheet, disrupting the desired smooth temperature distribution.
During the heating phase, high temperatures cause the large sheets to lose their structural integrity and sag downwards under their weight; see Fig. \ref{fig: sag}. This sagging leads to different gaps between heaters in the heat bank and their associated zones during the heating phase. The proposed method must be robust to sensitive system parameters, maintaining optimal control performance despite external disturbances. This robustness is crucial for ensuring consistent process outcomes and minimizing the need for manual adjustments.

\subsubsection{Scalability}

Another aspect is the control system's scalability, especially in industrial settings where thermoforming processes may involve hundreds of heaters. The proposed solution can scale without significant modifications since it uses a linear model in the control step by solving a Quadratic Programming (QP) problem. This scalability ensures that the control strategy can be efficiently applied across different machines and setups, facilitating its adoption in diverse industrial environments. 

\subsection{Proposed MPC for Thermoforming System}

Unlike the physics-based model, which requires evaluating all system states, the obtained NARX model has a lower order since it was trained using only the system's input and output data. This control-oriented model, when integrated into MPC, can significantly reduce the computational load of the optimization problem while maintaining the MPC's robust performance. As a result, the system can generate more precise control inputs, effectively achieving the objectives outlined in the previous section. The model inputs encompass the electrical power of the heaters within the heater bank, $u$, and the outputs comprise the average temperature of each zone on the thermoplastic sheet, $\mathcal{T}$. Equation \eqref{equ: regressors-thermoforming} shows the regressor vector for the NARX model proposed for thermoforming systems. 

\begin{equation} \label{equ: regressors-thermoforming}
X_t =  \left[\begin{array}{l}
\mathcal{T}_{t-1:t-n} \\
u_{t-1:t-m}
\end{array}\right],
\end{equation}

where $\mathcal{T}_{t-1:t-n} \in \mathbb{R}^{Z \times n}$ and $u_{t-1:t-m} \in \mathbb{R}^{H \times m}$ are the past $n$ and $m$ output and inputs at discrete-time index $t$, respectively (see also equation \ref{equ:NARX-general1}). $Z$ is the total number of zones on the thermoplastics, and $H$ shows the number of heaters. The system's input (heater's power) and output (sheet's temperature) $u_{t-n}$ and $\mathcal{T}_{t-n}$ , respectively, at time step $t-n$ takes the following form:

\begin{equation} \label{equ: NARX-thermoforming_output}
\mathcal{T}_{t-n}=\left[\begin{array}{c}
{\mathcal{T}_{t-n}}^{(1)} \\
{\mathcal{T}_{t-n}}^{(2)} \\
\vdots \\
{\mathcal{T}_{t-n}}^{(Z)}
\end{array}\right], 
{u}_{t-n}=\left[\begin{array}{c}
{u_{t-n}}^{(1)} \\
{u_{t-n}}^{(2)} \\
\vdots \\
{u_{t-n}}^{(H)}
\end{array}\right].
\end{equation}

The range of values $u^{(1)},\cdots,u^{(H)}$ is from 0 to 500 W, indicating the minimum and maximum power levels that are applied to each heater, indicating input admissible set $\mathcal{U}$ in equation \ref{equ: MPC general}. Assuming the system's $6$-second sampling time, collecting the dataset from the experimental setup to construct the NARX model would take a while for $20000$ data samples. Moreover, using several thermoplastic sheets was impractical due to the necessity to span a large temperature range and the possibility of melting at higher temperatures. To address this problem, the validated physics-based model presented in \cite{hosseinionari2023development}, based on Section \ref{sec: Modeling} heat transfer framework, was used to gather the required dataset.

\section{Results} \label{Results}

This section includes the model evaluation, the robustness and performance of the proposed method using the simulation environment, and experimental implementation. In detail, a comparison is made between the output results of the NARX model and the physics-based simulator \cite{hosseinionari2023development}. Then, the proposed control method performance is evaluated using the physics-based simulator and compared to two state-of-the-art methods: MPC-guided Deep Reinforcement Learning (DRL) \cite{HOSSEINIONARI2024DRL} and Adaptive MPC (AMPC) \cite{HOSSEINIONARI2024DRL}. Next, the proposed method's robustness against three important system parameters, convection heat transfer coefficient ($h$), the gap between the heater bank and the thermoplastic sheet ($d$), and the radiation absorptivity of the thermoplastic sheet ($\alpha$) are examined. Finally, the lab-scale experimental thermoforming setup is used to verify the proposed control algorithm as a physical platform.

\subsection{Model Evaluation: Fitness and Whiteness tests}

\begin{figure*}[h]
    \centering
    \includegraphics[width=1\linewidth]{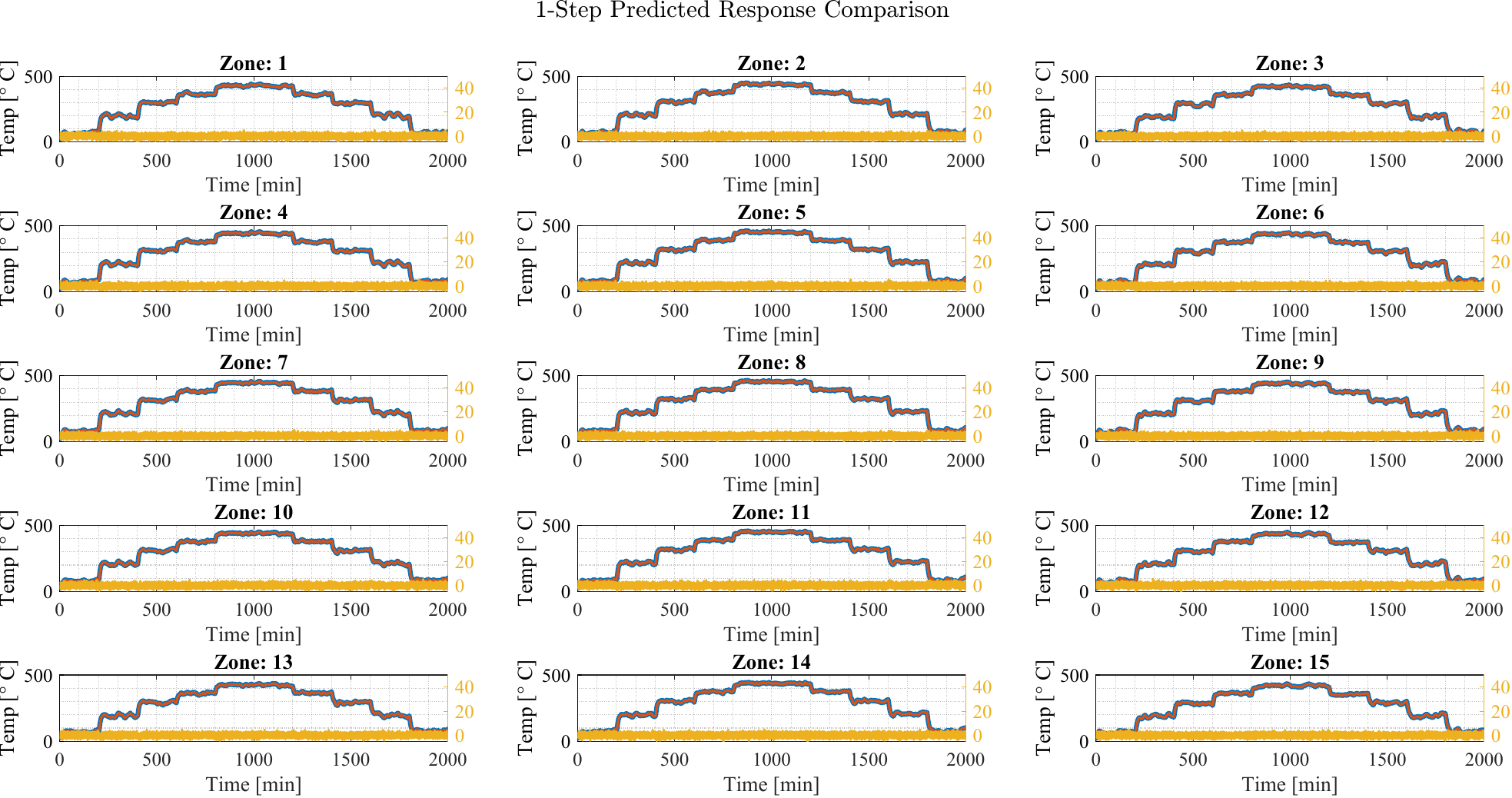}
    \caption{Comparison of 1-step temperature prediction and the actual temperature for 15 outputs (blue line: dataset, red line: NARX model's output, yellow line: error).}
    \label{fig: 1-step}
\end{figure*}

The accuracy of multi-step-ahead predictions can degrade due to the accumulation of one-step-ahead prediction errors during model recursion. Therefore, to assess the accuracy and robustness of the NARX model's multi-step-ahead predictions, N-step ahead predictions are also evaluated against the test set; see Table \ref{tab: closed-loop simulation prediction horizon}. It should be noted that the model's accuracy for the N-step prediction is critical as it will be used in the MPC design. The effectiveness of this experimental design is evaluated through the model's accuracy, quantified by the Normalized Root Mean Square Error (NRMSE) and presented in Fig. \ref{fig: 1-step}.
\begin{equation} \label{NRMSE}
\text{NRMSE} = \frac{\sqrt{\frac{1}{N_d} \sum_{i=1}^{N_d}(y_i - \hat{y}_i)^2}}{\max(y) - \min(y)},   
\end{equation}
where $N_d$ is the number of data points, $y$ is the noise-polluted measurement, and $\hat{y}$ is the model's predicted output. The blue and red lines display the dataset and model's predicted outputs in Fig \ref{fig: 1-step} for 1-step prediction. Additionally, the yellow line depicts the difference between these two values.

\begin{figure*}[h]
    \centering
    \includegraphics[width=1\linewidth]{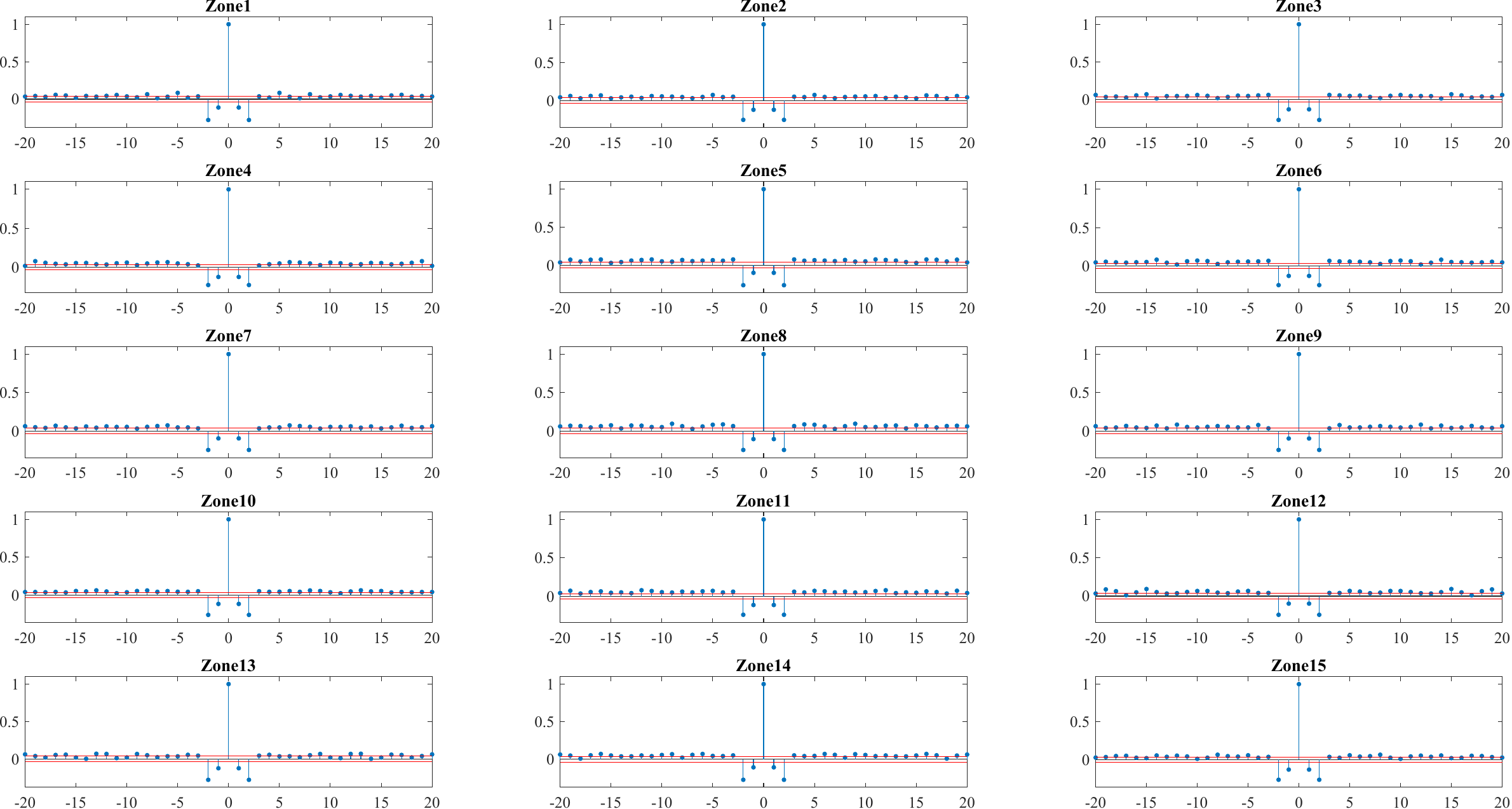}
    \caption{Sample NARX model's auto-correlation of residuals $e_j, j \in \{1,...,15\}$ with $99 \%$ confidence intervals: test dataset.}
    \label{fig: Auto-correlation}
\end{figure*}

\begin{figure*}[h]
    \centering
    \includegraphics[width=1\linewidth]{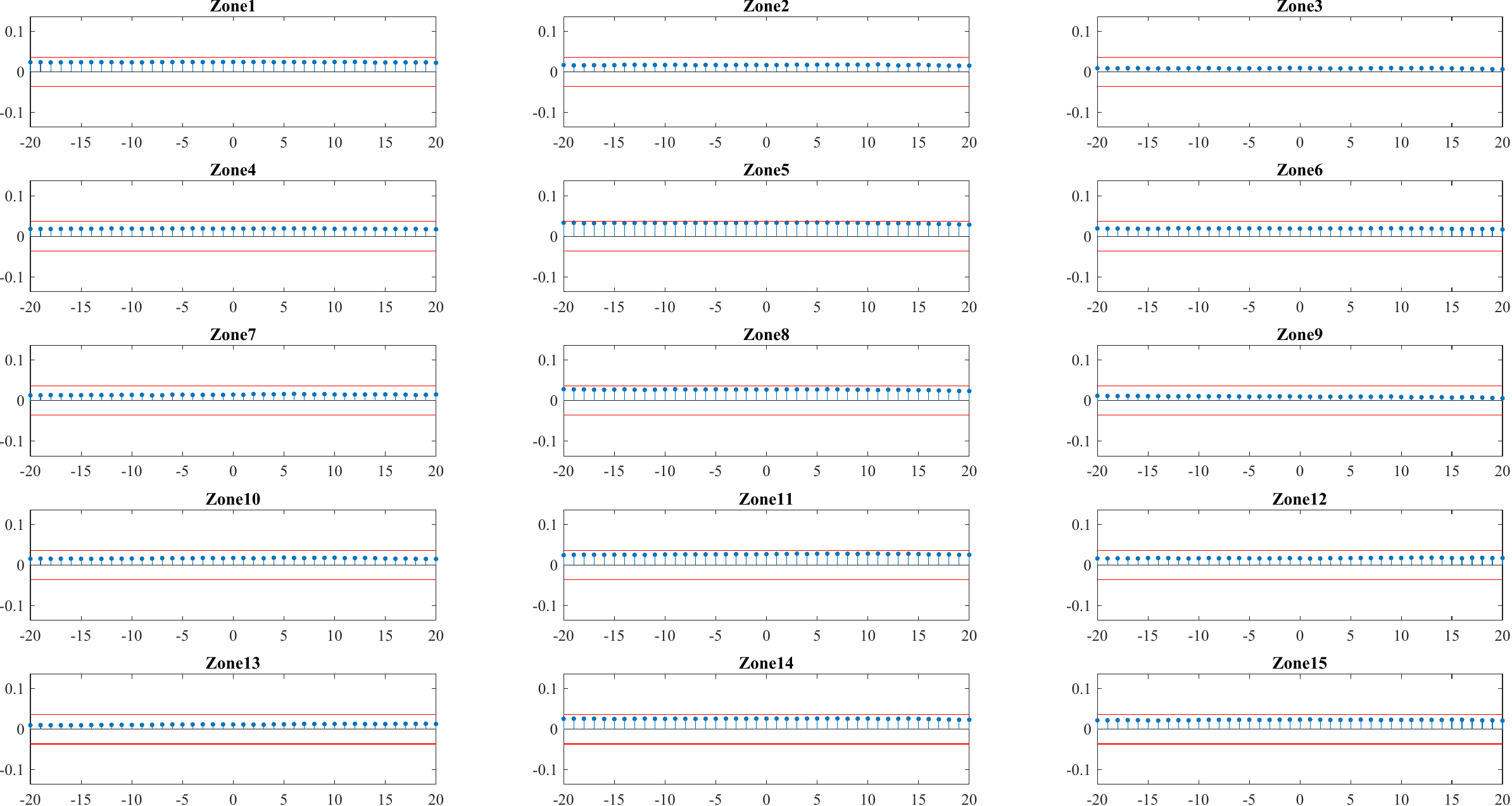}
    \caption{Sample NARX model's cross-correlation of input $u_j$ and residual $e_j, j \in \{1,...,15\}$ with $99 \%$ confidence intervals: test dataset.}
    \label{fig: Cross-correlation}
\end{figure*}

Figure \ref{fig: Auto-correlation} shows each zone's auto-correlation of residuals. Since most values lie in the $99 \%$ confidence intervals and there is a big spark on the center, we can conclude that the identified model has captured the essential dynamics, and the residual errors are close to the measurement noise. Sample cross-correlation in Fig. \ref{fig: Cross-correlation} indicates how the residuals are linearly dependent on lagged inputs or how well the proposed model has explored the information in the lagged inputs. It shows that the residuals and lagged inputs have little correlation as their cross-correlation lies in the $99 \%$ confidence interval. Some values in Fig. \ref{fig: Auto-correlation} and \ref{fig: Cross-correlation} are out of the $99 \%$ confidence intervals because the residuals contain both uncaptured nonlinear dynamics and measurement noise.

\begin{table}[!]
\centering
\caption{Accuracy of the proposed NARX model for the test data in the closed-loop simulation: NRMSE for each zone versus N-step prediction horizon (percentage).}
\label{tab: closed-loop simulation prediction horizon}
\begin{tabular}{|c|c|c|c|c|c|c|c|c|}
\hline
N & 1 & 10 & 30 & 50 & 80 & 100 \\ \hline
Zone 1 & 98.58 & 97.59 & 94.41 & 91.47 & 87.78 & 85.73 \\ \hline
Zone 2 & 98.59 & 97.46 & 93.97 & 90.74 & 86.68 & 84.44 \\ \hline
Zone 3 & 98.49 & 97.39 & 93.89 & 90.68 & 86.65 & 84.43 \\ \hline
Zone 4 & 98.61 & 97.48 & 93.99 & 90.80 & 86.77 & 84.54 \\ \hline
Zone 5 & 98.61 & 97.34 & 93.51 & 90.02 & 85.62 & 83.19 \\ \hline
Zone 6 & 98.57 & 97.32 & 93.52 & 90.05 & 85.67 & 83.24 \\ \hline
Zone 7 & 98.60 & 97.39 & 93.77 & 90.47 & 86.34 & 84.06 \\ \hline
Zone 8 & 98.62 & 97.24 & 93.22 & 89.57 & 84.99 & 82.46 \\ \hline
Zone 9 & 98.55 & 97.25 & 93.35 & 89.77 & 85.26 & 82.75 \\ \hline
Zone 10 & 98.58 & 97.44 & 93.92 & 90.68 & 86.60 & 84.35 \\ \hline
Zone 11 & 98.60 & 97.28 & 93.33 & 89.74 & 85.24 & 82.76 \\ \hline
Zone 12 & 98.58 & 97.34 & 93.61 & 90.21 & 85.93 & 83.56 \\ \hline
Zone 13 & 98.51 & 97.49 & 94.23 & 91.21 & 87.42 & 85.34 \\ \hline
Zone 14 & 98.52 & 97.33 & 93.62 & 90.21 & 85.94 & 83.59 \\ \hline
Zone 15 & 98.49 & 97.43 & 94.06 & 90.98 & 87.12 & 85.01 \\ \hline
\end{tabular}
\end{table}

Table \ref{tab: closed-loop simulation prediction horizon} presents the model's accuracy for each zone relative to the prediction horizon for test data, which was the final $25 \%$ of the entire data set. However, we optimized the NARX model for least one-step prediction error; it still needs to be evaluated for multi-step prediction. This is because, in MPC, we rely on the model's accuracy given a long prediction horizon. As it is clear, for the prediction horizon of $100$ steps, the model's accuracy is more than $80 \%$ on the test dataset.



\subsection{Controller Evaluation: Numerical Results}

To evaluate the controller's performance in various settings and conduct a comparative analysis with AMPC and MPC-guided DRL, the same base scenario introduced in \cite{HOSSEINIONARI2024DRL} is considered. Namely, Table \ref{tab: ref temp} shows the reference temperature distribution used in the comparisons. 

\subsubsection{Comparison Study}

\begin{figure*}[h]
    \centering
    \begin{subfigure}[b]{0.49\textwidth}
        \centering
        \includegraphics[width=\textwidth]{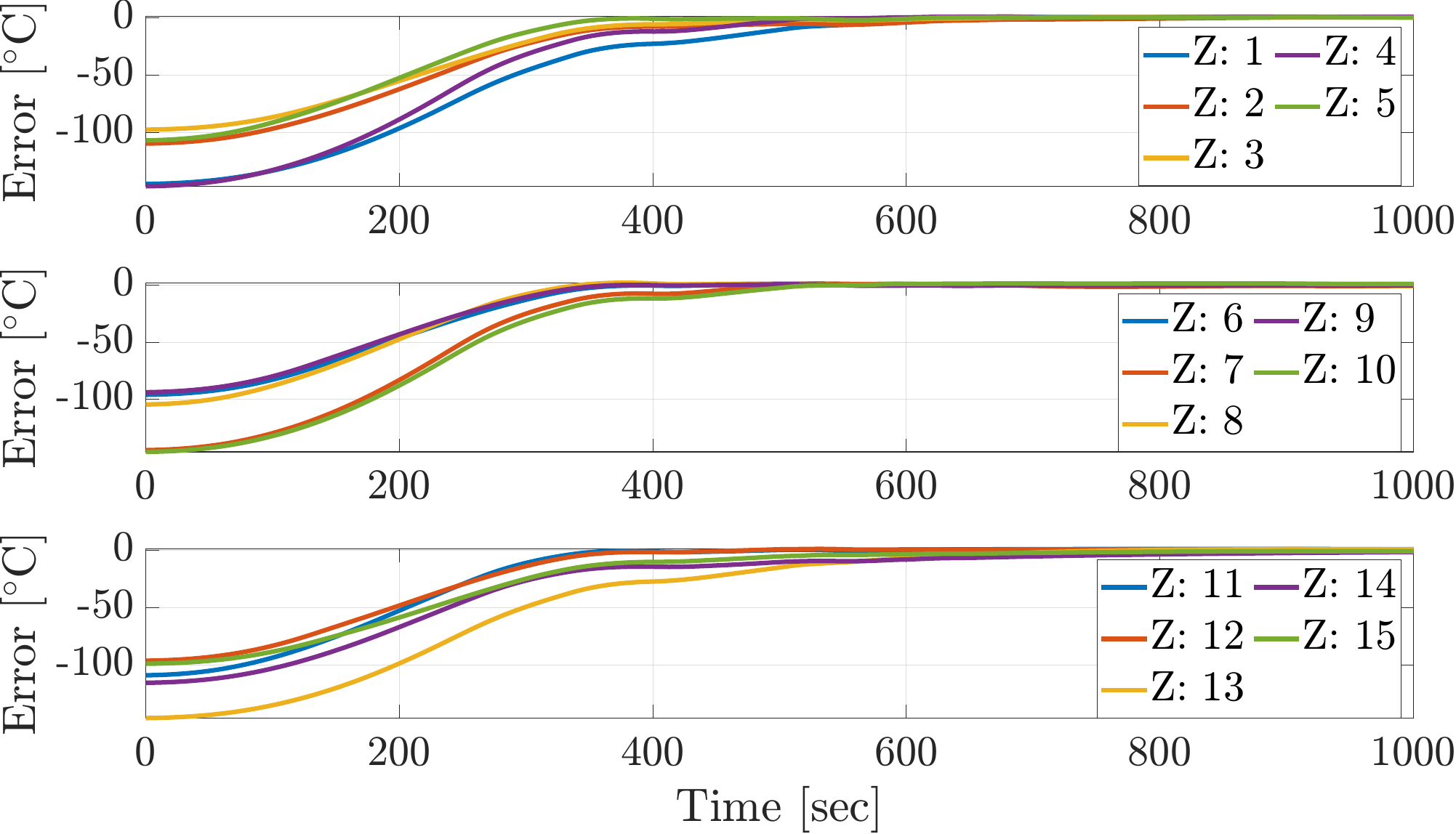}
        \caption{}
        \label{fig: simulator-nonuniform-zone error}
    \end{subfigure}    
    \begin{subfigure}[b]{0.49\textwidth}
        \centering
        \includegraphics[width=\textwidth]{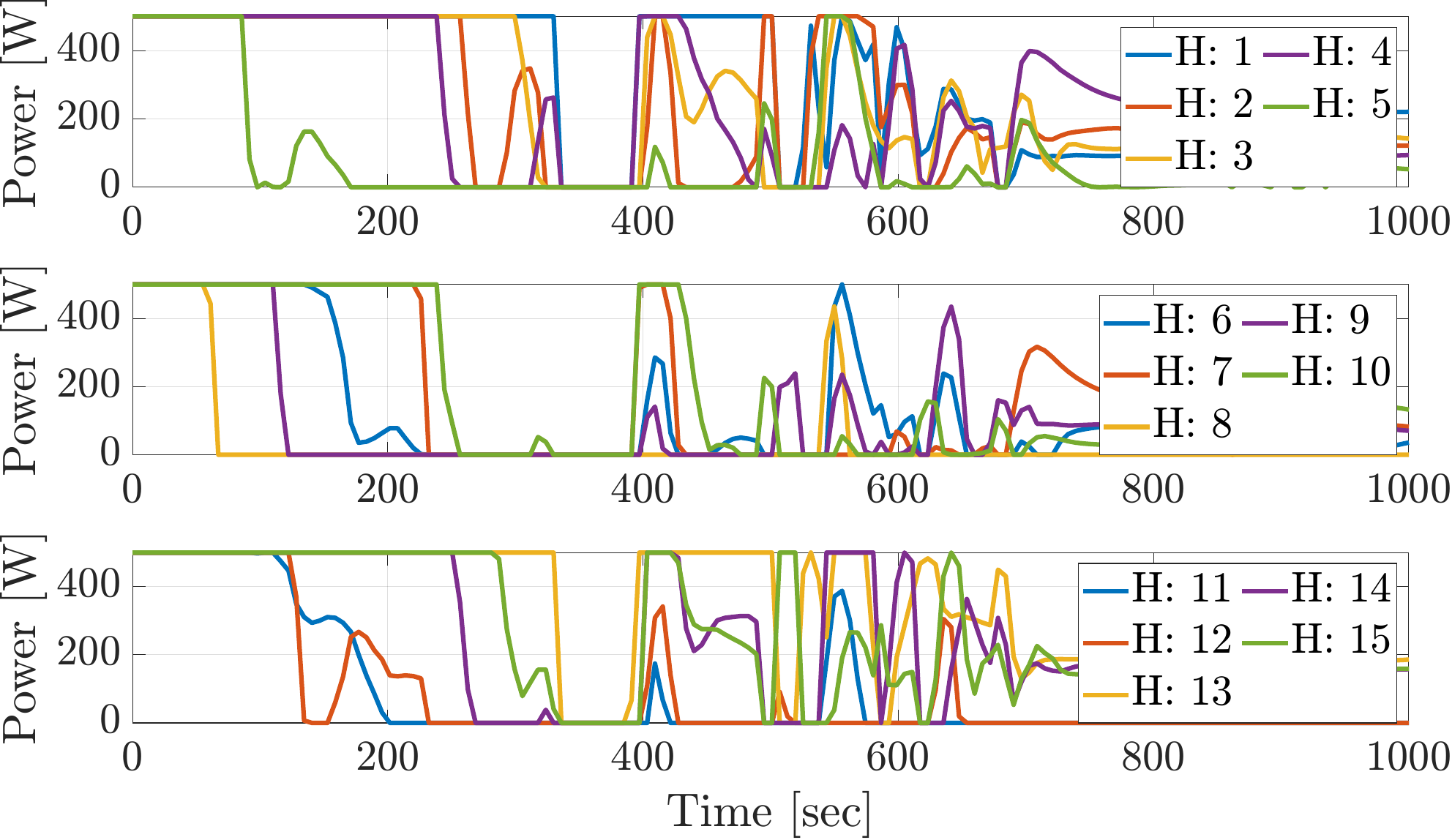}
        \caption{}
        \label{fig: simulator-nonuniform-u}
    \end{subfigure}    
    \caption{Simulation results of the proposed method applied on the physics-based simulator  ($H$ denotes each heater, and $Z$ denotes each zone). (a) error signals for the nonuniform reference temperature distribution. (b) input signals for the nonuniform reference temperature distribution.}
    \label{fig: simulator-nonuniform}
    
\end{figure*}

Several performance metrics for comparison are considered, including the average and the maximum error of zones at the end of the simulation time, the maximum overshoot, and the settling time within an error bound of $\pm10^\circ \mathrm{C}$. Figure \ref{fig: simulator-nonuniform-zone error} shows the error variation across the 15 zones on the thermoplastic sheet over the simulation time. After 560 seconds, the error signals lie in the $\pm10^\circ \mathrm{C}$ bound. The maximum overshoot and the average (maximum) error of zones are $2^\circ \mathrm{C}$ and $0.7^\circ \mathrm{C}$ ($1.4^\circ \mathrm{C}$), respectively. Also, Fig. \ref{fig: simulator-nonuniform-u} indicates the control inputs applied to 15 heaters bounded in $[0,500]$ Watts.

\newcolumntype{P}[1]{>{\centering\arraybackslash}p{#1}}
\begin{table}[!]
  \caption{Desired temperature distribution on the simulator and experimental implementation.}
    \begin{center}
        \begin{tabular}{P{0.55cm} P{0.55cm} P{0.55cm} P{0.55cm} P{0.55cm} P{0.55cm} P{0.55cm} P{0.55cm} P{0.55cm}}
        \hline
        \textbf{Zone}  & \textbf{Ref. Temp. Sim. [$^\circ \mathrm{C}$]} & \textbf{Ref. Temp. Exp. [$^\circ \mathrm{C}$]} &  \textbf{Zone}  & \textbf{Ref. Temp. Sim. [$^\circ \mathrm{C}$]} & \textbf{Ref. Temp. Exp. [$^\circ \mathrm{C}$]} &  \textbf{Zone}  & \textbf{Ref. Temp. Sim. [$^\circ \mathrm{C}$]} & \textbf{Ref. Temp. Exp. [$^\circ \mathrm{C}$]} \\
        \hline
 1 & 165.5 & 83 & 6  & 117   & 55 & 11 & 129.8 & 71 \\
 2 & 130.3 & 66 & 7  & 165.4 & 85 & 12 & 117.1 & 54 \\ 
 3 & 118.2 & 51 & 8  & 125.5 & 75 & 13 & 167   & 83 \\ 
 4 & 167.5 & 84 & 9  & 114.8 & 56 & 14 & 136.3 & 66 \\
 5 & 127.6 & 70 & 10 & 166.9 & 85 & 15 & 119.7 & 53\\
      \hline
      \end{tabular}
    \label{tab: ref temp}
    \end{center}
\end{table}

Table \ref{tab: comparision} compares the performance of the proposed method with AMPC and MPC-guided DRL across seven criteria. AMPC requires feedback from the surface temperature of heaters since it uses a linearized physics-based system model around operational points. In contrast, MPC-guided DRL and the proposed method are developed based on input (heaters' heating power) and output (zone temperatures) measurements. With an increase in the heaters in the heater bank, the system's inputs and outputs also increase, leading to higher complexity in the models utilized by the centralized MPC methods and, consequently, longer online computational time. However, while this complexity affects the offline training time of MPC-guided DRL, its online computational time remains low once trained, enabling fast operation. Regarding accuracy, the proposed method outperforms other methods, exhibiting a smaller final error bound. Specifically, the error bound achieved with the proposed method is $3.5$ times narrower than AMPC and $5.5$ times narrower than MPC-guided DRL. Utilizing the proposed method substantially reduces overshoot, decreasing from $12^\circ \mathrm{C}$ to $2^\circ \mathrm{C}$. This represents a significant six-fold reduction in overshoot. Furthermore, concerning settling time within an error bound of $\pm10^\circ \mathrm{C}$, the proposed MPC can achieve this error bound in 560 seconds. In contrast, the settling times for the AMPC and MPC-guided DRL methods are 650 seconds and 765 seconds, respectively.

\newcolumntype{P}[1]{>{\centering\arraybackslash}p{#1}}
\begin{table}[h]
    \centering
    \small
    \caption{A comparison between AMPC, MPC-guided DRL, and proposed MPC using linearized NARX model on the physics-based simulator.}
    \label{tab: comparision}
    \begin{tabular}{P{3cm} P{1cm} P{1.5cm} P{2cm} }
    \hline
    \textbf{Metric} & \textbf{AMPC} & \textbf{MPC-guided DRL} &  \textbf{Proposed MPC with Linearized NARX Model} \\
    \hline
    Reliance on model  & D\&I & Training & D\&I\\
    Requires heater temp. bank's feedback & Yes & No & No\\
    Average online comp. time [ms] & 921 & 11 & 90\\
    Average error after 1000 sec [$^\circ \mathrm{C}$] & 1.5 & 1.8 & \textbf{0.7}\\
    Maximum error after 1000 sec [$^\circ \mathrm{C}$] & $\pm$5 & $\pm$8 & $\pm$\textbf{1.4}\\
    Maximum overshoot [$^\circ \mathrm{C}$] & 12.0 & 11.8 & \textbf{2.0} \\
    Settling time [sec] & 650 & 765 & \textbf{560}\\
    \hline
    \multicolumn{2}{c}{D: Design I: Implementation}
    \end{tabular}
    
\end{table}

\subsubsection{Robustness Analysis}
This section assesses the robustness of the proposed method by examining its sensitivity to three critical system parameters: convection heat transfer coefficient ($h$), the gap between the heater bank and the thermoplastic sheet ($d$), and the radiation absorptivity of the thermoplastic sheet ($\alpha$). This robustness analysis enables us to assess the performance and stability of the proposed method before implementing it on the real-world setup in section \ref{Controller Evaluation: Experimental Results}. The proposed method's performance is evaluated based on the maximum overshoot and final average error across zones for different mentioned parameters. The nominal condition is defined as $h = 5 [\frac{W}{m^2}]$, $d = 15 [\mathrm{cm}]$, and $\alpha = 0.8$. 

\begin{figure*}[t]
     \centering
     \begin{subfigure}{0.225\textwidth}
         \includegraphics[width=\textwidth]{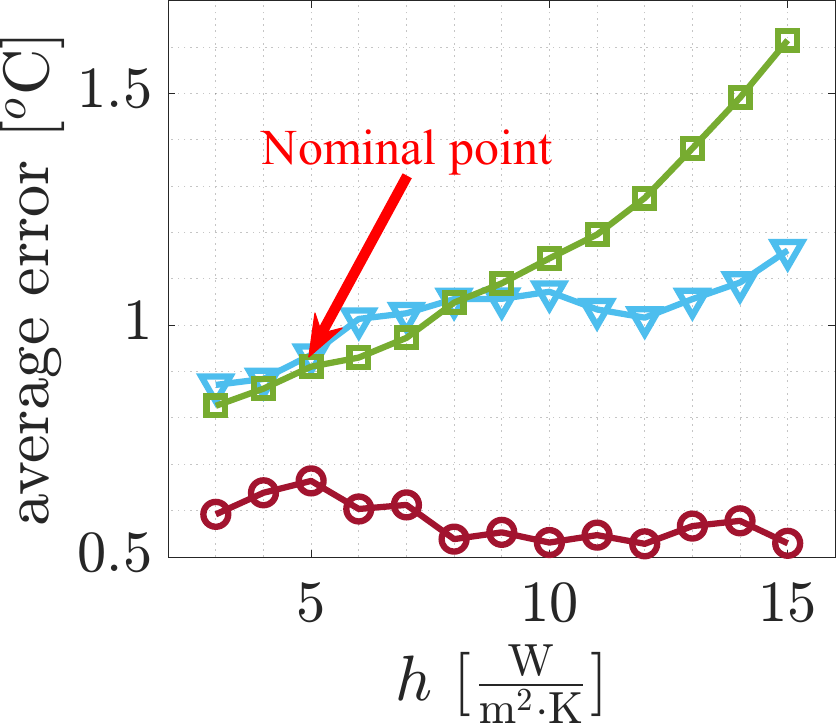}
         \caption{}
         \label{fig: sensivity1}
     \end{subfigure}
     \begin{subfigure}{0.23\textwidth}
         \includegraphics[width=\textwidth]{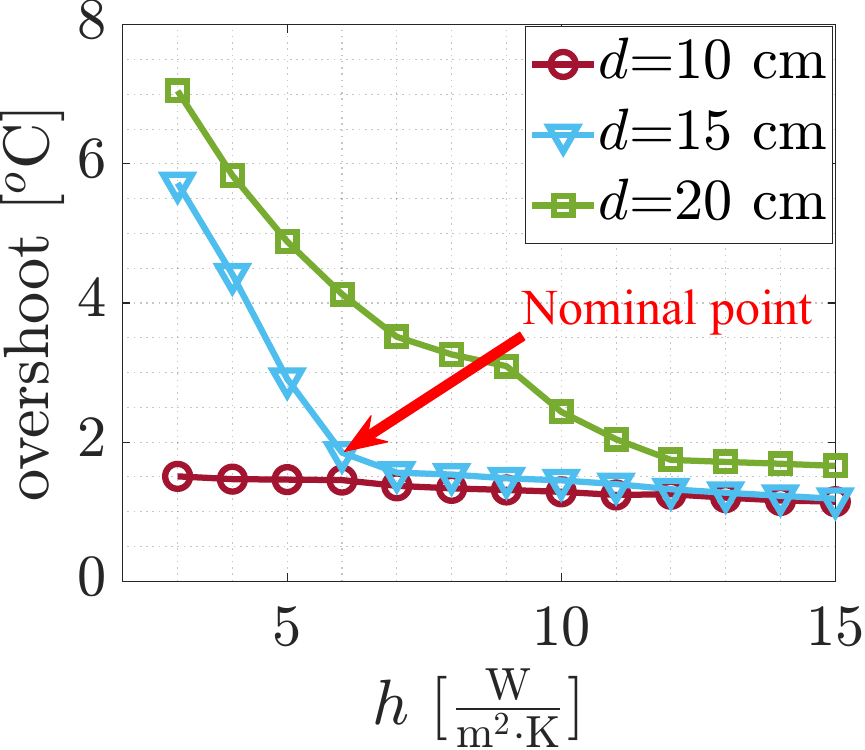}
         \caption{}
         \label{fig: sensivity2}
     \end{subfigure}
          \begin{subfigure}{0.225\textwidth}
         \includegraphics[width=\textwidth]{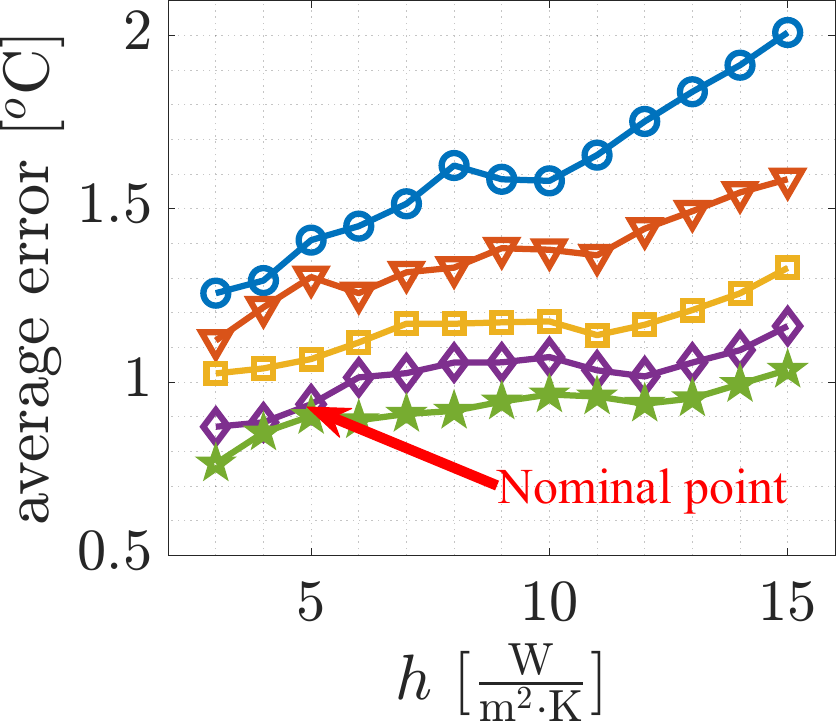}
         \caption{}
         \label{fig: sensivity3}
     \end{subfigure}
          \begin{subfigure}{0.23\textwidth}
         \includegraphics[width=\textwidth]{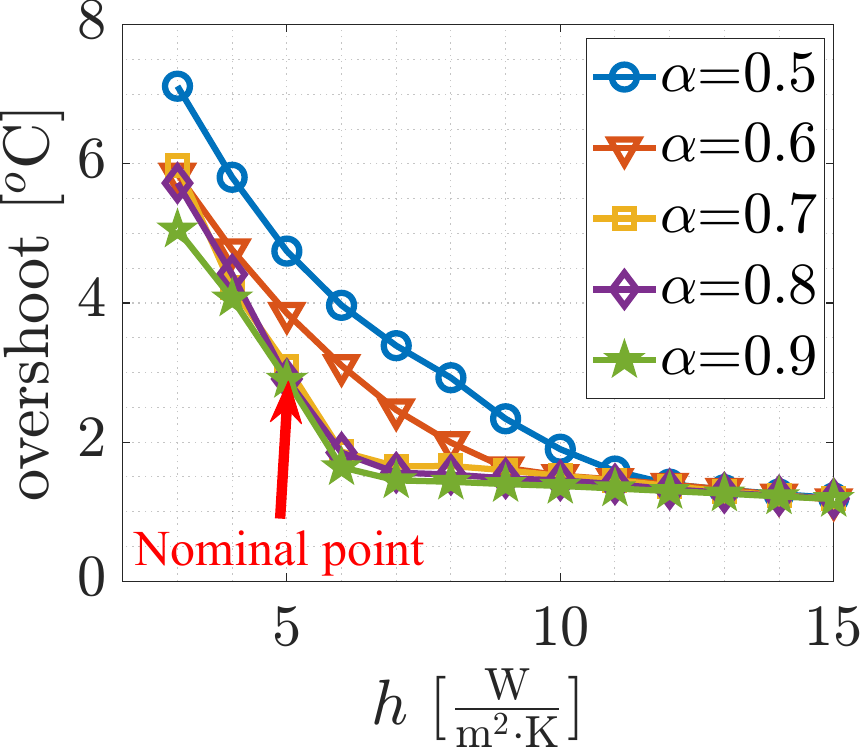}
         \caption{}
         \label{fig: sensivity4}
     \end{subfigure}
     \caption{Robustness evaluation: (a) The average error of zones under varying convection heat transfer coefficients and gaps between the heater bank and the sheet. (b) Maximum overshoot under varying convection heat transfer coefficients and gaps between the heater bank and the sheet. (c) The average error of zones under varying convection heat transfer coefficients and sheet absorptivity coefficients. (d) Maximum overshoot under varying convection heat transfer coefficients and sheet absorptivity coefficients.}
        \label{fig: sensivity}
\end{figure*}

In Figure \ref{fig: sensivity1} and Figure \ref{fig: sensivity2}, the heat transfer coefficient was varied for different distances between the sheet and the heaters while keeping the absorptivity at 0.8. In Figure \ref{fig: sensivity3} and Figure \ref{fig: sensivity4}, the heat transfer coefficient was varied for different absorptivity coefficients while keeping the distance between the heater bank and the sheet at 15 cm. Figure \ref{fig: sensivity1} and Figure \ref{fig: sensivity3} collectively indicate that despite significant variations in parameters, the performance of the proposed MPC remains robust. It consistently maintains the average error of zones on the sheet below 2 \textdegree C after the control process. Furthermore, as illustrated in Figure \ref{fig: sensivity2}, an increase in the distance between the heater bank and the sheet results in a larger overshoot. This outcome is logical, given that the heater's radiation pattern covers a broader area on the sheet, potentially impacting other zones as a disturbance. Additionally, Figure \ref{fig: sensivity} demonstrates that an elevation in the convection heat transfer coefficient (due to a drop in room temperature or exposure of the sheet to wind) leads to a decrease in both maximum overshoot and the final accuracy of the agent. The outcomes depicted in Figure \ref{fig: sensivity} indicate the proposed MPC is robust in handling disturbances introduced by changes in system parameters. These results instill confidence in the potential effectiveness of the trained multi-agent DRL when deployed in a real-world lab-scale setup.

\begin{figure*}[t]
    \centering
    \begin{subfigure}[b]{0.3\textwidth}
        \centering
        \includegraphics[width=\textwidth]{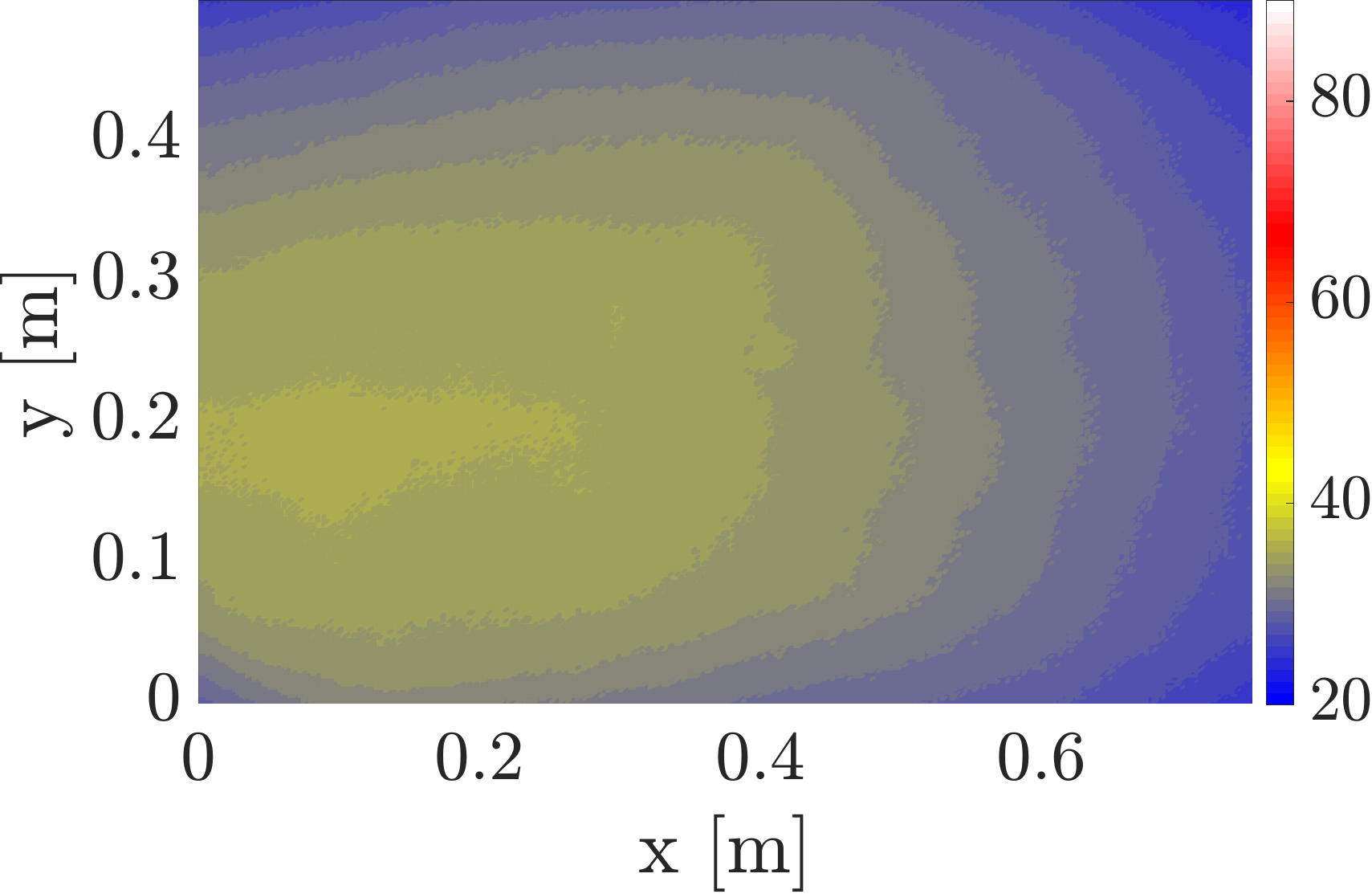}
        \caption{Temp. distribution at $t = 100 \, [sec]$}
        \label{fig:subfig_a}
    \end{subfigure}
    \hfill
    \begin{subfigure}[b]{0.3\textwidth}
        \centering
        \includegraphics[width=\textwidth]{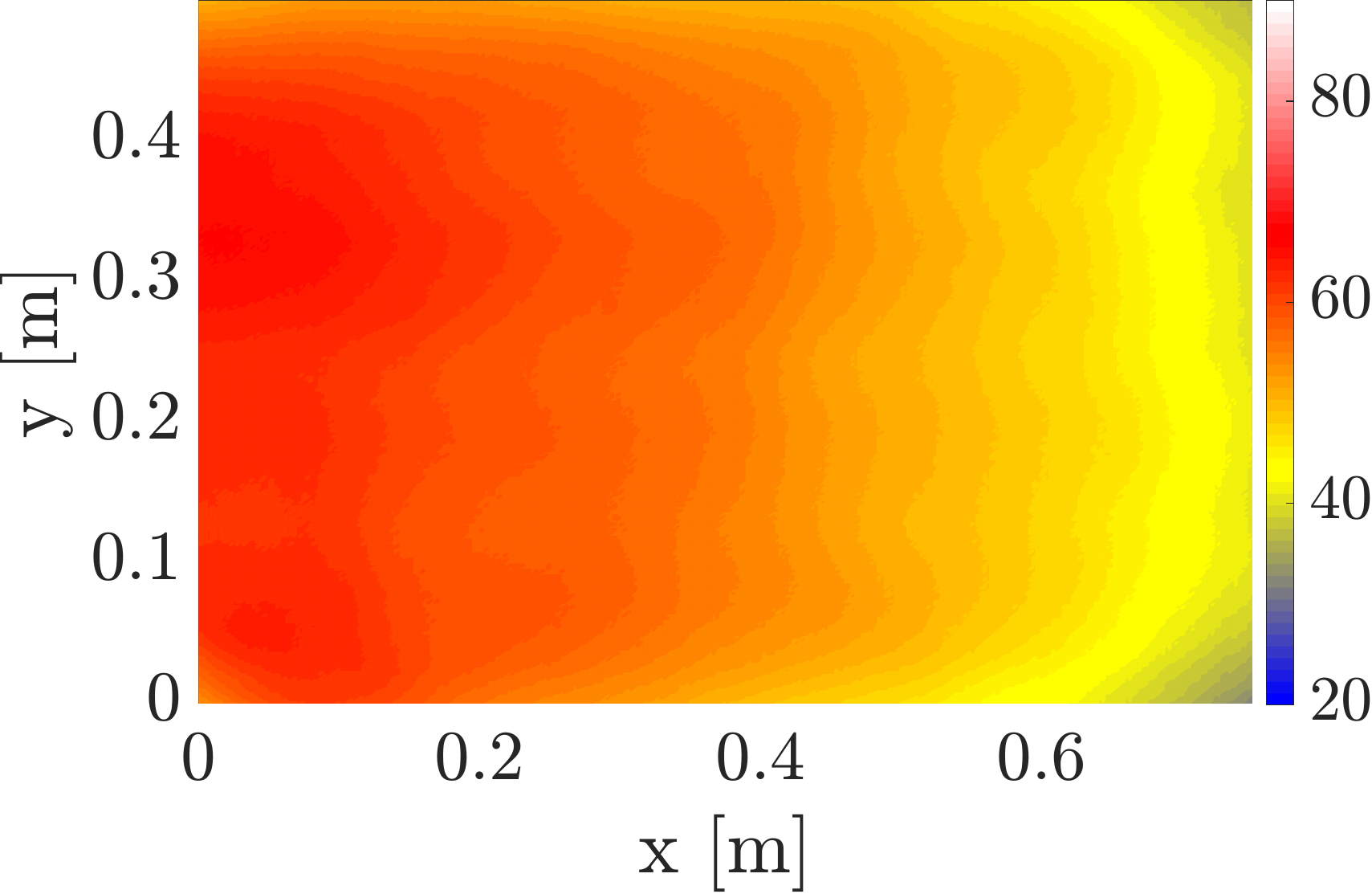}
        \caption{Temp. distribution at $t = 200 \, [sec]$}
        \label{fig:subfig_b}
    \end{subfigure}
    \hfill
    \begin{subfigure}[b]{0.3\textwidth}
        \centering
        \includegraphics[width=\textwidth]{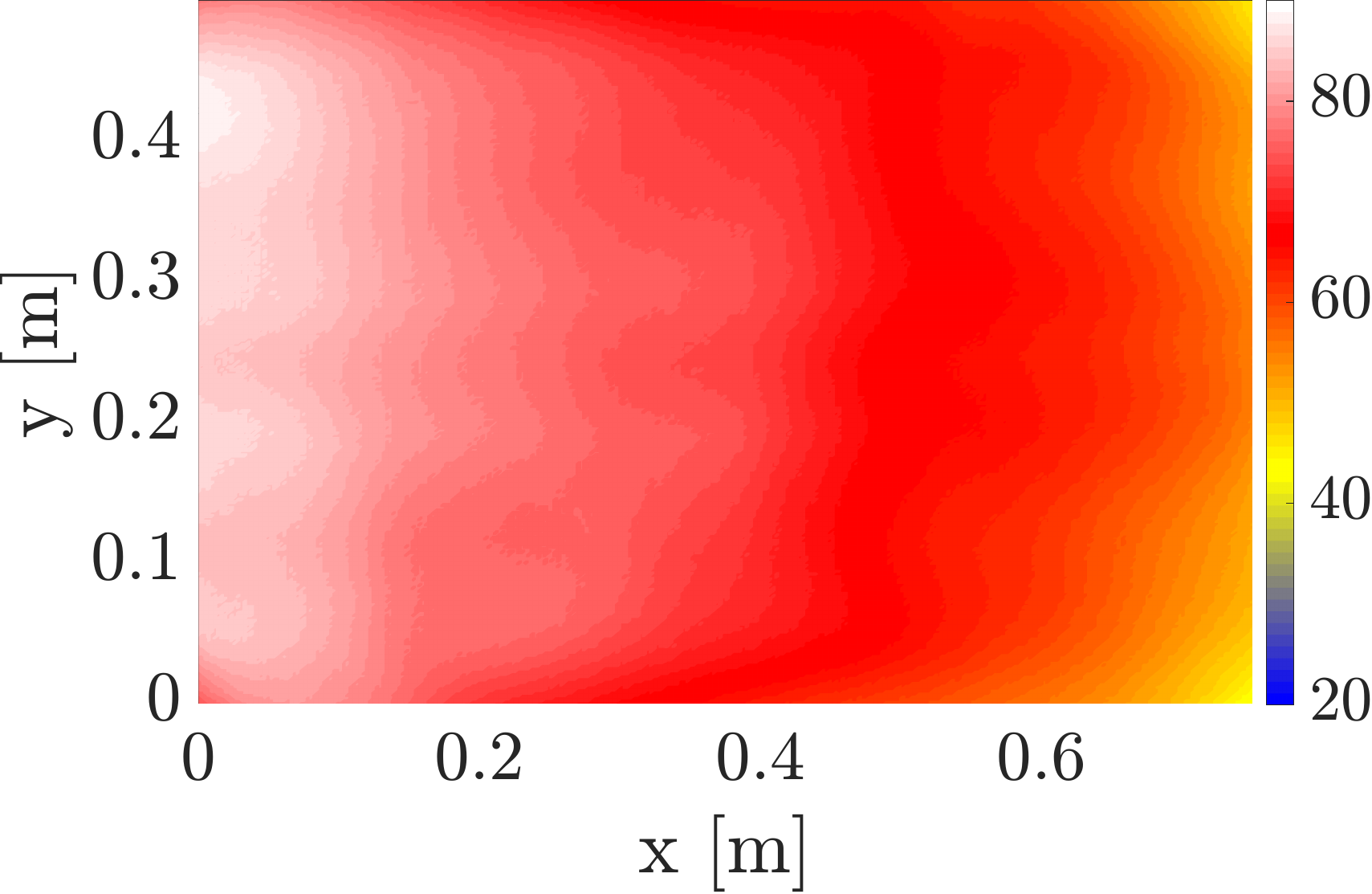}
        \caption{Temp. distribution at $t = 900 \, [sec]$}
        \label{fig:subfig_c}
    \end{subfigure}
    \vspace{0.5cm}
    \begin{subfigure}[b]{0.49\textwidth}
        \centering
        \includegraphics[width=\textwidth]{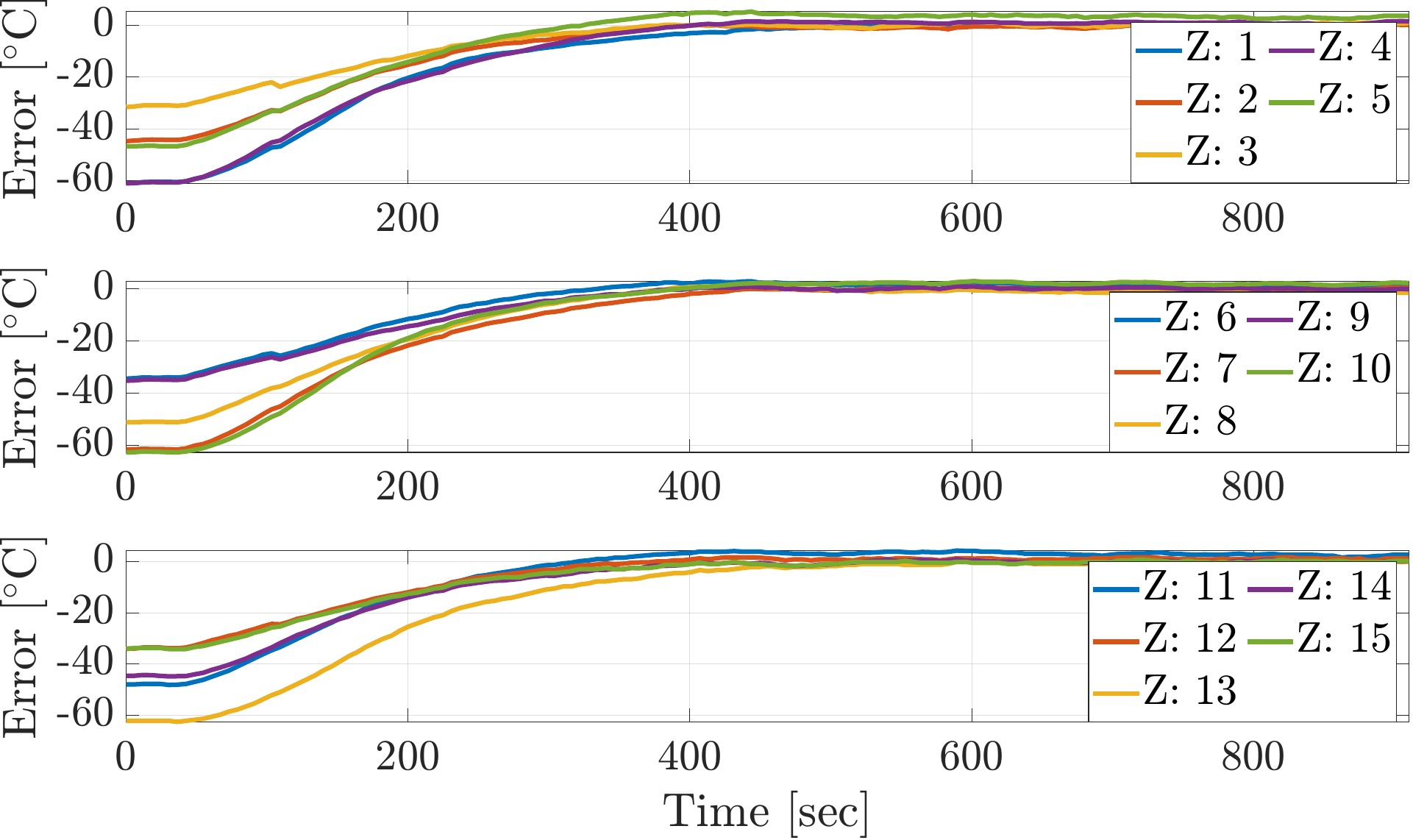}
        \caption{Error signals for the nonuniform reference temp. distribution.}
        \label{fig: experimental-nonuniform-zone error}
    \end{subfigure}    
    \begin{subfigure}[b]{0.49\textwidth}
        \centering
        \includegraphics[width=\textwidth]{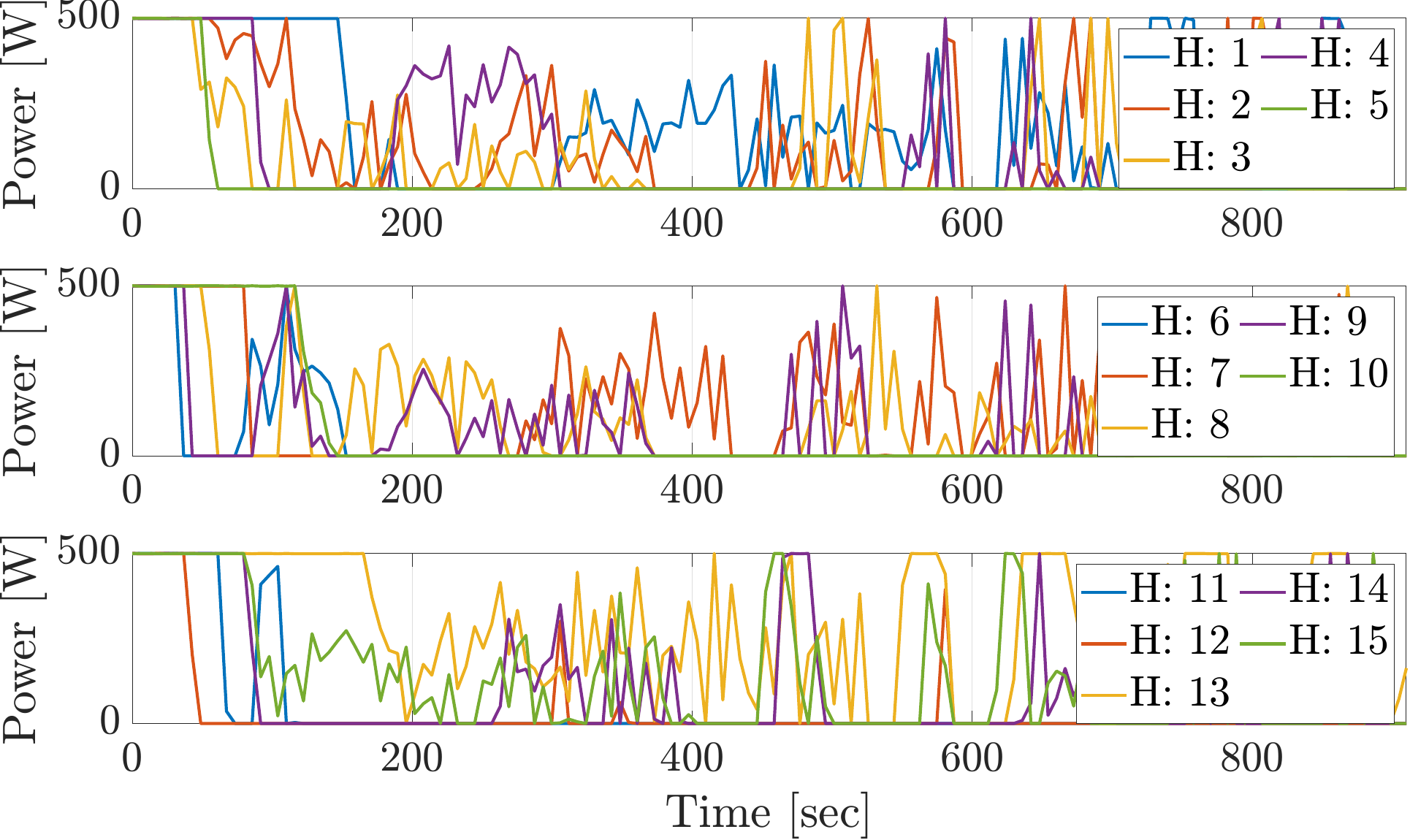}
        \caption{Input signals for the nonuniform reference temp. distribution.}
        \label{fig: experimental-nonuniform-u}
    \end{subfigure}    
    \caption{Experimental results of the proposed method applied on the lab-scale thermoforming system: measured by the IR camera.}
    \label{fig: experimental temperature profile}
    
\end{figure*}

\subsection{Controller Evaluation: Experimental Results} \label{Controller Evaluation: Experimental Results}

The practical application of the proposed method is evaluated on the experimental thermoforming setup, detailed in section \ref{sec: Experimental Setup}. A feasible non-uniform reference temperature distribution is chosen to evaluate the controller's performance, as shown in table \ref{tab: ref temp}, given the ambient temperature during the experimental test equal to $21^\circ \mathrm{C}$. Adopting the controller designed in the simulation step on the real setup, Fig. \ref{fig: experimental temperature profile} illustrates the system's output, measured temperature profile on the thermoplastic sheet by IR camera, at time stamps $t=100 [\mathrm{sec}]$, $t=200 \, [\mathrm{sec}]$, and $t=900 \, [\mathrm{sec}]$. Figure \ref{fig: experimental-nonuniform-zone error} depicts the error variation across the 15 zones on the thermoplastic sheet. It shows that the controller effectively reduces the error of zones within the $\pm$10 bound after $308$ seconds, with the average and maximum error of zones at the control process's conclusion being approximately $1^\circ \mathrm{C}$ and $3.8^\circ \mathrm{C}$, respectively, and the maximum overshoot equal to $5.3^\circ \mathrm{C}$. Additionally, Fig. \ref{fig: experimental-nonuniform-u} indicates the control inputs applied to the 15 heating elements.

Table \ref{tab: comparison sim vs. Exp} compares the proposed MPC controller's performance for the same reference temperature distribution on a physics-based simulator and a real-world lab-scale thermoforming setup. The results indicate a close alignment in performance when subjected to an identical reference temperature distribution on both platforms. Notably, the settling time within the experiment's error bound of $\pm 10 ^\circ \mathrm{C}$ was recorded as 308 sec, shorter than the simulator's settling time. Also, the overshoot on the real-world setup is almost $5 ^\circ \mathrm{C}$ higher than the simulator. This disparity suggests differences in system parameters between the simulator and the experiment. For instance, variations may arise if the convection heat transfer coefficient in the lab is lower than that considered in the simulator or if the sheet's radiation absorptivity is higher than the simulated values. 

\newcolumntype{P}[1]{>{\centering\arraybackslash}p{#1}}
\begin{table}[H]
    \centering
    \small
    \caption{Overall comparison of the simulation and experimental results for the non-uniform reference temperature distribution.}
    \label{tab: comparison sim vs. Exp}
    \begin{tabular}{P{4cm} P{1.5cm} P{1.5cm} }
    \hline
    \textbf{Metric} & \textbf{Simulator} & \textbf{Experimental} \\
    \hline
    Zones' average error [$^\circ \mathrm{C}$] & 0.6 & 1 \\
    Zones' maximum error [$^\circ \mathrm{C}$] & $\pm$2.6 & $\pm$3.8\\
    Overshoot [$^\circ \mathrm{C}$] & 0.5 & 5.3 \\
    Settling time (sec) & 380 & 308\\
    \hline
    \end{tabular}
\end{table}

\section{Conclusion}
\label{sec: Conclusion}

This work proposed an indirect data-driven predictive-based control method for thermal control in the thermoforming process. It outperformed state-of-the-art methods while benefiting from well-established tools and theories of system identification and predictive control. Several simulation studies were conducted using a high-fidelity simulator to evaluate the proposed method's robustness and performance under parametric uncertainty. The results indicated an overshoot and average steady-state error of less than $2^\circ \mathrm{C}$ and $0.7^\circ \mathrm{C}$ for the nominal scenario, and $7^\circ \mathrm{C}$ and $2^\circ \mathrm{C}$ for the worst-case scenario. Moreover, the experimental results of the proposed controller closely matched the simulation analysis, with overshoot and average steady-state error metrics of less than $5.3^\circ \mathrm{C}$ and $1^\circ \mathrm{C}$, respectively.

Furthermore, various simulation and experimental results indicated its applicability to industrial settings besides its low computational cost. By integrating a data-driven NARX model within the linear MPC framework, we address the critical challenges associated with traditional control methods, which often struggle with thermoforming systems' dynamic and nonlinear nature.

The experimental and simulation results confirmed that the proposed method significantly enhances overheating, operation time, and steady-state errors. Furthermore, our study assessed the proposed method's robustness to the system's sensitive parameters. Our approach's scalability and robustness make it a promising solution for thermoforming setups. The system's predictive capabilities allow for aggressive yet safe control actions to achieve faster response times. 

Future research will focus on direct data-driven control techniques that eliminate the SysID step and design the controller directly from the dataset. This may enable us to achieve better performance while simplifying the design process.

\bibliographystyle{IEEEtran} 
\bibliography{References}

\end{document}